\documentclass{article}
\usepackage[nosort]{cite}
\usepackage[margin=1.8in,footskip=0.25in]{geometry}
\usepackage{amssymb,amsmath,amsthm}
\usepackage{bbm}
\usepackage{bm}
\usepackage{multirow,multicol}
\usepackage{algorithm}
\usepackage{algorithmic}
\usepackage{graphicx}
\usepackage{xcolor}
\usepackage{adjustbox}
\usepackage[normalem]{ulem}
\usepackage{chngcntr}

\def\V{\mathcal{V}}

\def\S{\mathcal{S}}

\def\bbR{\mathbb{R}}

\newcommand{\norm}[1]{\lVert#1\rVert}

\newtheorem{proposition}{Proposition}

\newtheorem{theorem}{Theorem}

\usepackage{subcaption}

\begin{document}
\title{Probabilistic Structure Learning for EEG/MEG Source Imaging with Hierarchical Graph Prior	\thanks{P.~L.~Purdon is partially supported by NIH grants R01AG054081 and R01AG056015. Y.~Lou is partially supported by  NSF Awards DMS-1522786 and CAREER 1846690. R.~Li is supported in part by NSF grants  CCF-1527104 and DMS-1719620. \textit{Corresponding authors: L.~Wang, P.~L.~Purdon.}}}

\author{
	Feng~Liu$^{*}$,
	Li~Wang$^{*}$$^{\dag}$,~
	Yifei~Lou,~% <-this % stops a space
	Ren-Cang~Li,~
	Patrick~L.~Purdon\thanks{$*$ denotes equal contribution. F.~Liu and P.~L.~Purdon are with Department of Anesthesia, Critical Care and Pain Medicine, Massachusetts General Hospital, Harvard Medical School, Boston, MA 02119 USA and also with Picower Institute for Learning and Memory, MIT, Cambridge, MA, 02139 USA. L.~Wang and R.~Li are with the Department of Mathematics at The University of Texas at Arlington, Arlington, TX, 76013 USA. Y.~Lou is with the Department of Mathematical Sciences at The University of Texas at Dallas, Richardson, TX, 75080 USA. $\dag$ denotes corresponding author contribution.  Corresponding authors: L.~Wang (Email: li.wang@uta.edu) and P.~L.~Purdon (Email: patrickp@nmr.mgh.harvard.edu).}
}

\maketitle
\begin{abstract}
	Brain source imaging is an important method for noninvasively characterizing brain activity using Electroencephalogram (EEG) or Magnetoencephalography (MEG) recordings. Traditional EEG/MEG Source Imaging (ESI) methods usually assume that either source activity at different time points is unrelated, or that similar spatiotemporal patterns exist across an entire study period.  The former assumption makes ESI analyses sensitive to noise, while the latter renders ESI analyses unable to account for time-varying patterns of activity. To effectively deal with noise while maintaining flexibility and continuity among brain activation patterns, we propose a novel probabilistic ESI model based on a hierarchical graph prior. Under our method, a spanning tree constraint ensures that activity patterns have spatiotemporal continuity. An efficient algorithm based on alternating convex search is presented to solve the proposed model and is provably convergent.  Comprehensive numerical studies using synthetic data on a real brain model are conducted under different levels of signal-to-noise ratio (SNR) from both sensor and source spaces. We also examine the EEG/MEG data in a real application, in which our ESI reconstructions are neurologically plausible. All the results demonstrate significant improvements of the proposed algorithm over the benchmark methods in terms of source localization performance, especially at high noise levels.
\end{abstract}

\section{Introduction} \label{sec:intro}
Human brain is composed of roughly 100 billion neurons and brain functions are carried out by complex firing and interactions among the neurons, accompanied with electromagnetic, hemodynamic, and metabolic changes \cite{he2018electrophysiological}. As the electromagnetic is directly related to the neural firing activities, it reflects the real-time dynamical process of the brain, which can be directly measured by Electroencephalogram (EEG) and Magnetoencephalography (MEG). Both EEG and MEG yield a much higher temporal resolution up to a few milliseconds than other brain imaging modalities such as functional magnetic resonance imaging (fMRI), positron emission tomography (PET), and single-photon emission computed tomography (SPECT) \cite{he2018electrophysiological,babadi2014subspace,custo2014eeg,li2016epileptogenic}. However, one limitation of EEG/MEG is the low spatial resolution, as the corresponding measurements are acquired on the scalp with little information regarding neural activations inside the brain. Reconstructing a brain source signal from EEG/MEG measurements is known as EEG/MEG source localization or EEG/MEG source imaging (ESI) \cite{michel2004eeg}. The ESI techniques have been used in several clinical and/or brain research applications such as the study of language mechanisms, cognition process and sensory function with a brain-computer interface \cite{edelman2016eeg}, the localization of primary sensory cortex in evoked potentials for surgical candidates~\cite{willemse2016magnetoencephalographic}, and the localization of the irritative zone in focal epilepsy~\cite{megevand2014electric}~\cite{erem2017dynamic}.

%localizing  (13–15) or 

%ESI technique has been widely used in the study of language mechanisms, cognition process, sensory function, brain-computer interface \cite{edelman2016eeg} as well as the localization of epileptic seizure 

In general, the number of EEG/MEG sensors is much less than the number of brain sources and hence the ESI problem is highly ill-posed. In order to find a reasonable solution, it is necessary to impose certain neurophysiologically plausible assumptions as regularizations  \cite{michel2004eeg} \cite{grech2008review}. One seminal work to enforce a unique solution is by imposing the $\ell_2$-norm, referred to as minimum norm estimate (MNE)  \cite{hamalainen1994interpreting}, and its variants such as dynamic statistical parametric mapping (dSPM) method \cite{dale2000dynamic} and standardized low-resolution brain electromagnetic tomography (sLORETA)  \cite{pascual2002standardized}. The $\ell_2$-norm based methods usually lead to an over-diffuse solution. To overcome this drawback, Uutela et al. \cite{uutela1999visualization} adopted the $\ell_1$-norm, known as minimum current estimate (MCE), for sparse source reconstruction in ESI. 

However, the aforementioned approaches can potentially lead to inconsistency in the temporal direction, as source activities are estimated independently at each time point. To ensure temporal smoothness, a number of methods have been proposed. Mixed norm estimate (MxNE) \cite{gramfort2012mixed} uses an $\ell_{p,2}$-norm regularization  ($p\le 1$), which imposes $\ell_2$ on the time  course direction and $\ell_p$ ($p\le 1$) in the source space to ensure temporal smoothness and spatial sparsity. Later, irMxNE~\cite{strohmeier2016iterative} based on an iterative reweighted optimization scheme to solve the MxNE surrogate problem was proposed for source reconstruction solution and achieved better precision and stability. 
%\textcolor{green}{} {\color{blue}is about} time-frequency mixed-norm estimate (TF-MxNE) that recovers the sparse coefficients for the time-frequency dictionary for \textcolor{green}{better estimations} of non-stationary and transient source signals.
Gramfort \textit{et al.} proposed a time-frequency mixed-norm estimate (TF-MxNE) algorithm that recovers the sparse coefficients for the time-frequency dictionary for a better estimation of non-stationary and transient source signals~\cite{gramfort2013time}.  
%Also, structured sparse priors were considered in the time-frequency domain for better estimation of  non-stationary and transient source signals, hence the name of  time-frequency mixed-norm estimate (TF-MxNE) \cite{gramfort2013time}. 
A spatio-temporal unifying tomography (STOUT) algorithm \cite{castano2015solving} combined TF-MxNE in the time-frequency domain and sparse basis field expansions in the spatial space to get a temporal-spatial smooth solution.   Recently,  a  supervised method \cite{liu2017graphTBD} was proposed to achieve spatial sparsity and temporal consistency within class using label information derived graph regularization and $\ell_1$ norm.

In this paper, we are particularly interested in seeking a precise solution for ESI problem with the presence of noise in both sensor (channel) and source spaces. 
Due to the ubiquitousness of noise signal, the reconstructed source solution usually contains spurious sources that explain the measured noise. To properly deal with noise, certain pattern, such as classification/clustering structure, could be reasonably assumed for guiding precise source reconstruction. For example, the source activation pattern can form several representative micro-states if they are evoked by multiple exterior stimuli. In \cite{liu2017graphTBD}, the micro-states are assumed to be known as classification labels. However, the classification labels are generally difficult to acquire accurately. Moreover, it might be infeasible to find a one-to-one matching between the underlying source activation pattern with the correct label information when the underlying source signal follows a smoothing manifold structure with uncertain transition between brain states.

To overcome the above challenging issues, we propose a novel probabilistic model with a specially designed hierarchical prior that can effectively model both the unknown micro-states and manifold structure for ESI. Specifically, a set of landmarks are introduced to characterize the brain source signal. These landmarks can be considered as the unknown micro-states for characterizing the source activation patterns. Moreover, the unknown manifold pattern is further represented by a graph structure, where the vertexes are the landmarks. By modeling both the landmarks and the graph structure in a hierarchical prior, we can find effective estimates including reconstructed source signal, landmarks and the graph structure via the probabilistic hierarchical model.

The contributions of this paper can be summarized as follows: (1) We formulate a novel probabilistic model for ESI, in which the denoised micro-states and the manifold structure of the source activation pattern are presented as a hierarchical prior. (2) An efficient optimization method is proposed to solve the proposed model. The convergence of our proposed algorithm is provably guaranteed. (3) Extensive experiments on both simulated data and real data show that our proposed method outperforms the benchmark methods and demonstrates better results in the case of high level noise.

The remaining of this paper is organized as follows. In Section ~\ref{sec:model}, we  describe a probabilistic hierarchical model for ESI with three building blocks. We then present an efficient algorithm with guaranteed convergence in Section~\ref{sect:ours}. Both synthetic and real data are examined in Section~\ref{sect:exp} to demonstrate the performance of the proposed approach in comparison with the benchmark ones. Finally, conclusions are given in Section~\ref{sect:conclude}.

\section{Probabilistic Model with Hierarchical Prior} \label{sec:model}
We propose a novel probabilistic hierarchical model for the inverse problem of EEG/MEG data using priors. The model is specially designed based on three important ingredients: the sparsity of the source signal, the source signal denoising, and the relationship learning of the landmarks of source signals. In the following, we first present the probabilistic formulations of the three ingredients individually and then combine them to form the proposed hierarchical model.

\subsection{Sparsity prior in brain source imaging}

The EEG/MEG data measures the electromagnetic field at a set of $N_c$ sensors (or channels) for $N_t$ time points. We denote the EEG/MEG measurements as $X =[\mathbf{x}_1,\ldots,\mathbf{x}_{N_t}] \in \bbR^{N_c\times N_t}$. 
The linear mapping from the brain source to the sensors on the scalp is often referred to as the \textit{lead field} matrix obtained from the quasi-static approximation of Maxwell's equations \cite{wolters2004efficient}, denoted by $L \in \bbR^{N_c\times N_s}$, where $N_s$ is the number of distributed sources used to represent the discretized 3D head model. Specifically, each column of $L$ represents the electrical propagation weight vector from a particular source location to the EEG/MEG electrodes. Each row of of $L$ represents the  weights of contribution from all the sources to one channel. Given $X$ and $L$, the objective to find a source activation (source signal), denoted as $S = [\mathbf{s}_1, \ldots, \mathbf{s}_{N_t}]\in \bbR^{N_s\times N_t}$, where each column corresponds to electrical potentials in $N_s$ source locations for one of the $N_t$ time points.  The measured EEG/MEG data $X$ can be described as a linear function of sources $S$ with an additive noise,
\begin{align}
\mathbf{x}_i = L \mathbf{s}_i + \epsilon_i, \quad \forall i = 1, \cdots, N_t. \label{op:dl}
\end{align}
With proper whitening \cite{engemann2015automated}, the noise term $\epsilon_i$ can be assumed to follow the Gaussian distribution with mean zero and the  covariance identity matrix $I$, i.e., $\epsilon_i \sim \mathcal{N}(0, I)$. Therefore, the probability of $\mathbf x_i$ given $\mathbf s_i$ is expressed as
\begin{align}
p(\mathbf{x}_i | \mathbf{s}_i) \propto \mathcal{N}(L\mathbf{s}_i, I), \quad \forall i. \label{eq:likelihood}
\end{align}
Generally in EEG/MEG, the number of sources is much larger than the number of electrodes, i.e., $N_s \gg N_c$, which implies that  {(\ref{op:dl})} is an under-determined  {system} and results in an ill-posed problem of ESI. In order to properly recover $\mathbf s_i$ from $\textbf x_i$,  prior knowledge is often needed to formulate as a regularization term. A popular choice is   the $\ell_1$ penalty \cite{mairal2009online} that yields a sparse solution, which can be expressed by the Laplacian distribution. In particular, we assume each entry of $\mathbf{s}_i$ is i.i.d. drawn from a Laplacian distribution with mean zero and a positive parameter $\gamma_1$, i.e., 
\begin{align}
p(\mathbf{s}_i) = (\frac{\gamma_1}{2})^{N_s} \exp(-\gamma_1 \|\mathbf{s}_i\|_1). \label{eq:lap-prior}
\end{align}

To find source estimates with spatially sparse but smooth active regions, we introduce the spatial smoothness basis dictionary. The definition of spatial basis matrix can be found in \cite{haufe2011large}. By introducing the spatial basis function
\begin{equation}
\mathbf{x}_i = L \mathbf{s}_i + \epsilon_i = L \Psi \Psi^{-1}\mathbf{s}_i + \epsilon_i{,}
\label{eq:smoothness}
\end{equation}
where $\Psi = \left[\psi_1, \psi_2, \dots, \psi_{N_s}\right] $, and $\psi_{ij}$ is the $j$-th element in $\psi_i$, which is defined as:
\begin{align}
\psi_{ij}=\left\{
\begin{array}{lcl}
1       &      & {i =j }{,}\\
\exp(-\|d_{ij}/\varrho\|^2)    &      & {j \in \Omega_i}{,}\\
0     &      & {\mbox{otherwise,}}
\end{array} \right. 
\label{eq:kernelfun}
\end{align}
where $d_{ij}$ is the distance between {sources} $i$ and $j$, $\varrho$ is the spatial smoothness scalar controlling the smoothness level, and $\Omega_i$ is a set of nearest neighboring sources of $i$ based on geodesic distance. 
Equation (\ref{eq:smoothness}) can be written as  $\mathbf{x}_i = \tilde{L} \mathbf{\tilde{s}}_i + \epsilon_i$ with $\tilde{L}= L \Psi$ and $\mathbf{\tilde{s}}_i = \Psi^{-1}\mathbf{s}_i$. 
Without introducing confusion, we still use $L$ instead of $\tilde{L}$ and the final reconstructed signal will be left multiplied by $\Psi$ to get the source signal $S$.

\subsection{Source signal denoising prior over latent landmarks}

In the EEG/MEG data, noise is inevitable in both the sensor and source spaces. To effectively deal with various kinds of noise, 
we innovatively consider to denoise the source signals as a probabilistic density estimate problem. The denoised signals are then constrained by a graph structure, which will be detailed in Section~\ref{sect:graph}. 

We formulate the  source signal denoising  as hierarchical prior in a probabilistic framework. Intuitively, the source signal denosing becomes simple if we know the true distribution of the clean source signals. In addition to the sparse prior (\ref{eq:lap-prior}), we assume that the  true  distribution of clean source signals depends on a set of latent variables $C:=[\mathbf c_1,\cdots,\mathbf c_{K}]\in\bbR^{N_s\times K}$, which we call the landmarks for the ease of reference. Informally, the true distribution of the clean source signal $\mathbf{s}$ considered as a random variable is denoted by $p(\mathbf{s} | C )$. For any given noisy source signal $\mathbf{s}_i$, we can measure the difference between $p(\mathbf{s}_i | C )$ and the posterior distribution $p(\mathbf{s}_i | X )$ obtained by applying the Bayes rule on (\ref{op:dl}) and (\ref{eq:lap-prior}). Instead of directly calculating the difference of two distributions based on a known true distribution, we model the source signal denoising as a prior for ESI by further modeling the latent landmarks in a probabilistic hierarchical model. 

Given $K$ landmarks in $C$, we employ kernel density estimation (KDE) on $C$ and use the estimate for the approximation to the true distribution. The basic idea of KDE involves smoothing each point $\mathbf{s}_i$ by a  kernel function  and  summing up all these functions together to obtain a final density estimation. A typical choice of the kernel function
is a Gaussian function, defined as $\kappa(\mathbf{y}) = (2 \pi)^{D/2} \exp(-\frac{1}{2} \mathbf{y}^T \mathbf{y})$, where $D$ is the dimension of the input $\textbf y$. Applying KDE to estimate $\mathbf{s}_i$ over landmarks $C$ leads to the following density function,
\begin{align}
p(\mathbf{s}_i | C ) = \frac{1}{K \sigma^{N_s}} \sum_{k=1}^K (2 \pi)^{\frac{N_s} 2} \exp (-\frac{1}{2\sigma^2} ||\mathbf c_k - \mathbf{s}_i||^2 ), \label{eq:kde-prior}
\end{align} 
where $\sigma$ is the bandwidth parameter of the  Gaussian kernel function. According to the Gaussian kernel function, the landmark is the mean of denoised source signal $\mathbf{s}$, so it is reasonable for $\mathbf{c}$ to inherit the sparsity property of $\mathbf{s}$. As a result, we consider the landmarks to follow the same Laplacian distribution as $\mathbf s_i$ but with a different parameter $\gamma_2$, i.e.,
\begin{align}
p(\mathbf{c}_k) = (\frac{\gamma_2}{2})^{N_s} \exp(-\gamma_2 \|\mathbf{c}_k\|_1), \quad \forall k=1,\cdots, K. \label{eq:lap-prior-c}
\end{align} 
The denoising property of this prior  will be  further discussed in Section \ref{sec:hmodel}.

\subsection{Landmark prior via graph structure learning}\label{sect:graph}
Recall that the landmarks are introduced to model the true distribution of the source signals. Instead of assuming independence among these landmarks, we attempt to model the pairwise relationships between landmarks in terms of some constrained graphs.

We assume the landmarks have an underlying graph structure, denoted by $G \in \mathbb{R}^{K \times K}$, where each vertex of $G$ corresponds to one landmark. 
%Recall the landmarks are $C=[\mathbf c_1,\cdots,\mathbf c_{K}] \in \mathbb{R}^{N_s \times K}$. 
Denote $\mathbf f_l$ as the $l$th row vector of $C$, which is often referred to as the feature vector. We assume that $\mathbf f_l$ is i.i.d. drawn from the following distribution,
\begin{align}\label{eq:dist4feature}
p(\mathbf f_l | G) \propto \exp\left( - \frac{\beta}{2} \sum_{k=1}^K \sum_{k'=1}^K g_{k, k'} (f_{l,k} - f_{l,k'})^2 \right), 
\end{align}
where $\beta$ is a positive parameter. This form of distribution \eqref{eq:dist4feature} enforces smoothness among features \cite{zhu2003semi} in the sense that a larger $g_{k,k'}$ results in a smaller distance of $(f_{l,k} - f_{l,k'})^2$. Since each row of $C$ is assumed to be i.i.d., we obtain the joint distribution
\begin{align}
p(C | G) \propto& \exp\left( - \frac{\beta}{2} \sum_{k=1}^K \sum_{k'=1}^K g_{k, k'} \sum_{l=1}^{N_s} (f_{l,k} - f_{l,k'})^2 \right) \notag\\
=& \exp\left( - \frac{\beta}{2} \sum_{k=1}^K   \sum_{k'=1}^K g_{k, k'}\| \mathbf c_k - \mathbf c_{k'} \|^2 \right).\label{eq:graph}
\end{align}
In order to make $p(C|G)$ plausible as a probability measure,  $G$ is required to be non-negative and symmetric, i.e., $g_{k,k'} \geq 0$ and  $g_{k,k'}=g_{k',k}, \forall k, k'$, respectively. 

In general, $G$ is unknown, so it is challenging to model the density function over all graphs that have $C$ as their vertices. To make the structure learning pragmatic, constraints on the set of graphs are required. To overcome this challenge, we directly model $\log p(G)$ instead of modeling $p(G)$ for learning a specific type of graphs from data such as the spanning trees \cite{Cheung2008}. Informally, given a connected undirected graph $\mathcal{T}=(\mathcal{V},\mathcal{E})$ with an edge $(V_k,V_{k'}) \in \mathcal{E}, \forall k, k'$, let $\mathcal{T}$ be a tree and $\mathcal{E}$ be the edges forming a tree.
In order to represent and learn a tree structure, $\{ g_{k,k'} \}$ are formulated as binary variables where $g_{k,k'}=1$ if $(V_k,V_{k'}) \in \mathcal{E}$ and $0$ otherwise, i.e., $G=[g_{k,k'}] \in \{0,1\}^{K \times K}$. To this end, we express the parametric formulation of $\log p(G)$ as an indicator function of the set of trees given by
\begin{align}
\log p(G) \propto \left\{
\begin{array}{ll}
0, & G \in \mathcal{T}, \\
-\infty, & \textrm{otherwise}
\end{array}
\right., \label{eq:tree-prior}
\end{align}
where $\mathcal{T} = \{ G \in \{0,1\}^{K \times K}\} \cap \{G = G^T\} \cap \{ \frac{1}{2} \sum_{k,k'} g_{k,k'} = |\mathcal{V}| -1 \} \cap \{ \frac{1}{2} \sum_{ V_k \in \S, V_{k'} \in \S} g_{k,k'} \leq |\S|-1, \forall \S \subseteq \mathcal{V}  \}$ \cite{Cheung2008}. The third constraint in $\mathcal{T}$ states that the spanning tree only contains $|\V|-1$ edges and the fourth constraint imposes the acyclicity and connectivity properties of a tree. Equation (\ref{eq:tree-prior}) states that any graph as a tree is uniformly sampled from $p(G)$, and otherwise the probability is zero.

\subsection{The proposed probabilistic hierarchical model } \label{sec:hmodel}

Given the likelihood function (\ref{eq:likelihood}) and priors (\ref{eq:lap-prior}), (\ref{eq:kde-prior}), (\ref{eq:lap-prior-c}), and (\ref{eq:graph}), we are use the Bayes rule ready to formulate the joint conditional distribution $p(S, C, G| X, L)$, which is proportional to 
\begin{align}\label{eq:all_prob}
\prod_{i=1}^{N_t} \Big[ p(\mathbf{x}_i | \mathbf{s}_i)  p(\mathbf{s}_i) p(\mathbf{s}_i | C)\Big]  p(C | G) \Big[\prod_{k=1}^K p(\mathbf{c}_k) \Big] p(G).
\end{align}
To obtain estimates $S$, $C$ and $G$ from data, we propose to employ the maximum a posteriori estimation by minimizing the negative logarithm of the conditional probability $p(S, C, G| X, L)$, given by,
\begin{align}
\min\limits_{S,C,G} & \|X - LS\|^2_{F} +\gamma_1\sum\limits_{i=1}^{N_t}\|\mathbf s_i\|_1 + \gamma_2 \sum\limits_{k=1}^{K}\|\mathbf  c_k\|_1\notag - 2 \lambda \sigma^2 \sum_{i=1}^{N_t} \log \sum_{k=1}^K \exp(- \frac{\| \mathbf{s}_i - \mathbf  c_k \|^2}{2 \sigma^2}) \nonumber\\
& +\frac{\beta}{2}\sum\limits_{k=1}^{K}\sum\limits_{k'=1}^{K}\|\mathbf c_k-\mathbf c_{k'}\|^2g_{k,k'} - \log p(G). \label{op:joint}
\end{align}

\begin{figure}[t]
	\centering
	%	\begin{subfigure}[b]{0.49\textwidth} 
	\includegraphics[width=\linewidth,height=0.5\linewidth]{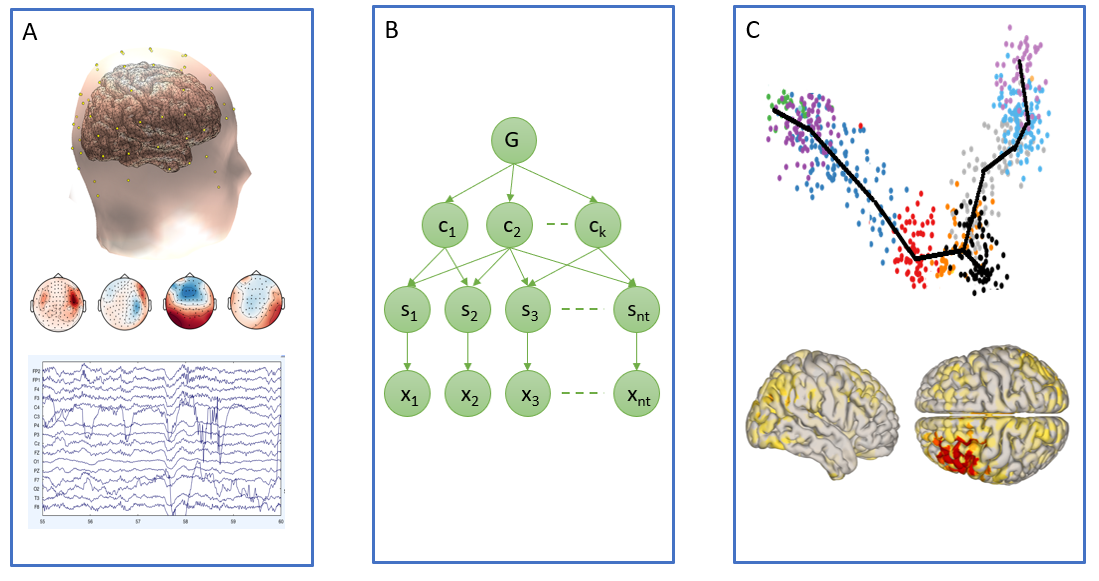}%
	\caption{Proposed ESI framework. A:  Building  forward model and  preprocessing  EEG/MEG data; B: ESI with hierarchical graph prior; C: Source imaging with reconstructed signal.} \label{fig:pipeline}
\end{figure}

The ESI framework is illustrated in Fig.\ref{fig:pipeline}. To better understand the proposed model (\ref{op:joint}), we explore its many properties as follows:
%take the further exploration on its properties:

\subsubsection{Graph structure learning via minimum-cost spanning tree} Problem (\ref{op:joint}) with respect to $G$, namely the graph structure learning, is equivalent to the minimum-cost spanning tree (MST) \cite{Cheung2008}. By  {substituting (\ref{eq:tree-prior}) into (\ref{op:joint})}, we have the following optimization problem with respect to $G$:
\begin{align}
\min_{G \in \mathcal{T}} \frac{\beta}{2}\sum\limits_{k=1}^{K}\sum\limits_{k'=1}^{K}\|\mathbf c_k-\mathbf c_{k'}\|^2g_{k,k'}, \label{op:MST-problem}
\end{align}
which is a problem of the minimum-cost spanning tree with the cost for the edge $(V_k, V_{k'})$ defined as $\frac{\beta}{2} \|\mathbf c_k-\mathbf c_{k'}\|^2$. 

%\vspace{-1em}
\subsubsection{Source signal denoising via landmarks} Based on the kernel density estimation \eqref{eq:kde-prior}, 
we obtain the probability of any point that is not in the set of landmarks. In particular, we can introduce an assignment matrix $R\in\bbR^{N_t\times K}$,  {whose} each entry $r_{i,k}$  {represents} the probability of assigning $\mathbf s_i$ to $\mathbf c_k$. We can obtain the following useful results \footnote{some proof is omitted in this version}: 

\begin{proposition}\label{prop:kde}
	Let $\alpha$ be a positive value and the feasible set $\mathcal{R} =\{r_{i,k}|r_{i,k}\geq 0,\sum\limits_{k=1}^ {K}r_{i,k}=1\,\forall i\}$ {. We have}
	{	\begin{align}
		\tilde{g}(S, C) = \min_{R \in \mathcal{R}} g(S, C, R), \label{eq:intro-R}
		\end{align}
		where two functions are defined as
		\begin{align}
		\tilde{g}(S, C) &=  -\alpha \sum_{i=1}^{N_t} \log \sum_{k=1}^K \exp(- \frac{\| \mathbf{s}_i - \mathbf{c}_k \|^2}{\alpha}), \\
		g(S, C, R) &=  \sum_{i=1}^{N_t} \sum\limits_{k=1}^{K}[r_{i,k}\|\mathbf{s}_i-\mathbf{c}_k\|^2+\alpha r_{i,k}\log r_{i,k}]. \label{eq:r-kde}
		\end{align}}
\end{proposition}

\begin{proposition} \label{prop:r} 
	Given $C$ and $S$,  {minimizing $g(S,C,R)$ with respect to $R\in \mathcal{R}$} has a closed form solution
	\begin{align}
	r_{i,k} = \frac{\exp\left(-\frac{\|\mathbf{s}_i-\mathbf{c}_k\|^2}{\alpha}\right)}{\sum\limits_{k=1}^{K}\exp\left(-\frac{\|\mathbf{s}_i-\mathbf{c}_k\|^2}{\alpha}\right)},\quad \forall i,k. \label{eq:r}
	\end{align}
\end{proposition}
%\begin{proof}
%	This is obtained directly according to the KKT conditions in the proof of Proposition \ref{prop:kde} {.}
%\end{proof}

\begin{proposition} \label{prop:s}   {Given $R \in \mathcal{R}$ and $C$, $\min\limits_{S} g(S,C,R)$ has the closed form solution}
	\begin{align}
	s_i = \frac{\sum_{k=1}^K r_{i,k} \mathbf{c}_k}{\sum_{k=1}^K r_{i,k}}, \forall i.
	\end{align}
\end{proposition}

Propositions \ref{prop:r} and \ref{prop:s} state that the source signal $S$ is automatically adjusted according to the weighting strategy $R$ and the latent landmarks $C$. Suppose that $\mathbf{s}_j$ is one noisy source signal in $S$ and the landmarks capture the high density region according to KDE. Proposition \ref{prop:r} tells us the weighting $r_{j, k}$ can be very small if $\mathbf{s}_j$ is far away from $\mathbf{c}_k$ because the distance between the source signal and the landmark can be very large. Proposition \ref{prop:s} implies that the source signal $\mathbf{s}_j$ will converge to the high density region of the most relevant landmarks (high weights). The representation of the noisy source signal to a robust point inside the high density region of landmarks is preferred for  {denoising source signals} since landmarks are the representative points of the true distribution of clean source signals. 

\section{The proposed optimization approach}\label{sect:ours}

According to Proposition \ref{prop:kde} and letting $\alpha = 2\sigma^2$, the estimates $S$, $C$ and $G$ of (\ref{op:joint}) can be equivalently obtained by introducing $R \in \mathcal{R}$ and solving the following problem 
\begin{align}
\min\limits_{S,C,G \in \mathcal{T},R \in \mathcal{R}} h(S,C,G, R),\label{op:joint-tree:model}
\end{align}
where $h(\cdot)$ is defined as
\begin{align}
h(S,C,G,R):=& \|X - LS\|^2_{F}+\frac{\beta}{2}\sum\limits_{k,k'=1}^{K} g_{k,k'} \|\mathbf c_k-\mathbf c_{k'}\|^2+\lambda\sum\limits_{i=1}^{N_t}\sum\limits_{k=1}^{K}[r_{i,k}\|\mathbf s_i-\mathbf c_k\|^2\notag\\
& +\alpha r_{i,k}\log r_{i,k}] +\gamma_1\sum\limits_{i=1}^{N_t}\|\mathbf s_i\|_1 + \gamma_2\sum\limits_{i=1}^{K}\|\mathbf c_k\|_1. \label{op:joint-tree}
\end{align}

We consider an alternating convex search (ACS) method \cite{gorski2007biconvex} to solve the proposed model  {(\ref{op:joint-tree:model})}. The ACS algorithm iterates as follows,
\begin{equation}\label{eq:ASC}
\left\{\begin{array}{l}
S^{(n+1)} =\arg\min\limits_{S} h(S,C^{(n)},G^{(n)}, R^{(n)}),\\
C^{(n+1)} = \arg\min\limits_C h(S^{(n+1)},C,G^{(n)}, R^{(n)}),\\
G^{(n+1)} = \arg\min\limits_{G\in\mathcal T} h(S^{(n+1)},C^{(n+1)},G, R^{(n)}),\\
R^{(n+1)} = \arg\min\limits_{R\in\mathcal R} h(S^{(n+1)},C^{(n+1)},G^{(n+1)}, R),
\end{array}
\right.
\end{equation}
where $n$  {is} the iteration number. We omit the iteration number when the context is clear. 
Below we will describe how to solve each subproblem in details. 

We rewrite the $S$-subproblem in \eqref{eq:ASC} as
\begin{align}\nonumber
%S: &= \arg\min\limits_{S} \|X - LS\|^2_{F}+\lambda\sum\limits_{i=1}^{N_t}\sum\limits_{k=1}^{K}r_{i,k}\|s_i-c_k\|^2+\gamma_1\sum\limits_{i=1}^{N_t}\|s_i\|_1, \\ \nonumber
& \arg\min\limits_{S} \sum\limits_{i=1}^{N_t} \Big( \|\mathbf{x}_i- L \mathbf{s}_i\|^2_{2}+\lambda \mathbf{s}_i^T \mathbf{s}_i -2 \lambda (\sum\limits_{k=1}^{K}r_{i,k} \mathbf{c}_k^T ) \mathbf{s}_i + \gamma_1 \|\mathbf{s}_i\|_1\Big) \nonumber\\
%& = \arg\min\limits_{S} \sum\limits_{i=1}^{N_t} \left( s_i^TL^TLs_i - 2x_i^TLs_i+\lambda s_i^Ts_i -2 \lambda \left(\sum\limits_{k=1}^{K}r_{i,k} c_k^T \right)s_i+\gamma_1 \|s_i\|_1\right), \\\nonumber
%& = \arg\min\limits_{S} \sum\limits_{i=1}^{N_t} \left( s_i^T(L^TL+\lambda I) s_i - 2\left(x_i^TL+\lambda\sum\limits_{k=1}^{K}r_{i,k} c_k^T \right)s_i+\gamma_1 \|s_i\|_1\right),\\\nonumber
%& = \arg\min\limits_{S} \sum\limits_{i=1}^{N_t} \left( s_i^T \tilde{L} s_i - 2b_i^T s_i+\gamma_1 \|s_i\|_1\right), \\\nonumber
& = \arg\min\limits_{S} \sum\limits_{i=1}^{N_t} \left( \mathbf{s}_i^T U^TU \mathbf{s}_i - 2 \mathbf{b}_i^T \mathbf{s}_i+\gamma_1 \|\mathbf{s}_i\|_1\right)  \nonumber\\ 
& = \arg\min\limits_{S} \sum\limits_{i=1}^{N_t}  \|U \mathbf{s}_i-U^{-T} \mathbf{b}_i\|_2^2+\gamma_1\| \mathbf{s}_i\|_1,  \label{S:Lasso:1} 
\end{align}
where $U$ is the  {Cholesky factor of $L^TL+\lambda I = U^TU$} and $\mathbf{b}_i = \big( \mathbf{x}_i^TL+\lambda\sum\limits_{k=1}^{K}r_{i,k} \mathbf{c}_k^T \big)^T = L^T \mathbf{x}_i+\lambda\sum\limits_{k=1}^{K}r_{i,k} \mathbf{c}_k $. It is straightforward that  $U$ is invertible. By letting $Y = U^{-T}[\mathbf{b}_1,\ldots,\mathbf{b}_{N_t}] = U^{-T}\left(L^TX+\lambda CR^T\right)$, we  {see} the $S$-subproblem is 
equivalent to solving $N_t$  {independently} strictly convex subproblem:
\begin{equation}\label{S:Lasso:sub:1}
\mathbf{s}_t :=  \arg\min\limits_{\mathbf{s}_t} \|U \mathbf{s}_t-U^{-T} \mathbf{b}_t\|_2^2+\gamma_1\|\mathbf{s}_t\|_1.
\end{equation}	

The $C$-subproblem in \eqref{eq:ASC}  {can be} expressed as
\begin{align}
\label{sub:C:1}
C^{(n+1)} =&\arg\min\limits_{C} \frac{\beta}{2}\sum\limits_{k,k'=1}^{K}\|\mathbf c_k-\mathbf c_{k'}\|^2g_{k,k'}\notag\\
&+\lambda\sum\limits_{i=1}^{N_t}\sum\limits_{k=1}^{K}r_{i,k}\|\mathbf s_i-\mathbf c_k\|^2+\gamma_2 \sum\limits_{k=1}^{K}\|\mathbf c_k\|_1.
\end{align}
Denote by $P= \text{diag}(G\mathbf{1})-G$ and $\Lambda =\text{diag}(\mathbf{1}^TR)$, where $\mathbf{1}$ is a vector of all ones. Note that $P$ is the graph Laplacian matrix defined on graph $G$. Let $\|C\|_{1,1} = \sum\limits_{k=1}^{K}\|\mathbf{c}_k\|_1$. We obtain an equivalent form of \eqref{sub:C:1} as follows,
\begin{equation}\label{sub:C:2}
\min\limits_{C}\text{trace}\left(C\left(\beta P+\lambda \Lambda \right)C^T-2\lambda SR C^T\right)+\gamma_2 \|C\|_{1,1}.
\end{equation}	
This is an (unconstrained) strictly convex problem, which admits a unique solution. To  {solve} \eqref{sub:C:2}, we consider the Cholesky decomposition of the matrix $\beta P+\lambda \Lambda:= VV^T$, where $V\in \bbR^{K\times K}$. Since the Laplacian matrix $P$ is positive semidefinite  and the diagonal matrix $\Lambda$ is positive definite ($r_{i,k}>0$) according to Proposition \ref{prop:r}, the matrix $\beta P+\lambda \Lambda$ is positive definite, which guarantees that $V$ is invertible. As a result, we have the solution $C$ given by: 
\begin{equation}\label{solution:C}
C := \arg\min\limits_{C}\|V^TC^T - \lambda V^{-1}R^TS^T\|_{F}^2+\gamma_2\|C\|_{1,1}.\\
\end{equation}
Both (\ref{S:Lasso:sub:1})  and (\ref{solution:C}) are $\ell_1$ regularized quadratic programming problem, which are  strictly convex and hence there exists a unique solution for each subproblem.    Furthermore,  it can be efficiently solved by many well developed methods such as Homotopy  \cite{asif2014sparse}, ADMM LASSO \cite{boyd2011distributed} and FISTA \cite{beck2009fast}.
%Various sparse algorithms are discussed in the survey \cite{zhang2015survey}. 
In this paper, we adopt the Homotopy solver for the above two problems. 

The $G$-subproblem in \eqref{eq:ASC} boils down to an MST problem as shown in problem (\ref{op:MST-problem}). It can be solved efficiently by the Kruskal's algorithm \cite{kruskal1956shortest}. In addition, for the $R$-subproblem in \eqref{eq:ASC}, there is a closed-form solution given by (\ref{eq:r}).

In summary, the overall ACS algorithm for solving the proposed model  {(\ref{op:joint-tree:model})} is given in Algorithm~\ref{alg} with the following initializations of variables: $S$ is solved by FISTA \cite{beck2009fast} for the inverse problem, $C$ is obtained by applying the $K$-means method {\cite{lloyd1982least}} on the initialized $S$, $G$ is the MST using the Kruskal's algorithm and $R$ is updated using (\ref{eq:r}). 
%By the strict convex property, if the optimal solution exists, it must be unique. 
The convergence of this algorithm is established in Theorem~\ref{thm:conv}. 

\begin{algorithm}[t]
	\caption{The proposed algorithm}\label{alg}
	\begin{algorithmic}
		\STATE Input: Data $X\in \mathbb{R}^{N_c\times N_t}$, matrix $L\in \mathbb{R}^{N_c\times N_s}$, parameters $\beta,\lambda,\alpha,\gamma_1,\gamma_2, K$.
		
		\textbf{Initialize $S$, $C$, $G$, $R$} 
		%		\STATE obtain $S$ by FISTA \cite{beck2009fast}
		%		\STATE obtain $C$ by applying the $K$-means method on $S$
		%		\STATE obtain $G$ by solving (\ref{op:MST-problem}) using Kruskal's algorithm
		%		\STATE obtain $R$ using (\ref{eq:r})
		\REPEAT
		\STATE update $S$ by solving (\ref{S:Lasso:sub:1})
		\STATE update $C$ using (\ref{solution:C})
		\STATE update $G$ by solving (\ref{op:MST-problem}) using Kruskal's algorithm
		\STATE update $R$ using (\ref{eq:r})
		
		\UNTIL{convergent.}
		\STATE Output: $S$, $C$, $G$, $R$
	\end{algorithmic}
\end{algorithm}
%\vspace{-1em}

\begin{theorem}\label{thm:conv}
	Suppose $\{S^{(n)},C^{(n)}, G^{(n)},R^{(n)}\}$  {is} the optimal solution of problem  {(\ref{op:joint-tree:model})} in the $n$-th iteration with each subproblem solved exactly. Let $h^{(n)}=h(S^{(n)},C^{(n)}, G^{(n)},R^{(n)})$ be the corresponding objective function value {. We} have
	\begin{enumerate}
		\item[(a)] 	 {the} function $h(S,C,G, R)$ is coercive {;}
		\item[(b)] $h^{(n+1)}\leq h^{(n)}$, i.e., objective function value is monotonically decreasing {;}
		\item[(c)]  {the} sequence $\{h^{(n)}\}$ converges as $n\rightarrow \infty$ {;}
		\item[(d)]  {the} sequence $\{S^{(n)},C^{(n)}, G^{(n)},R^{(n)}\}$ has a convergent subsequence.
	\end{enumerate}	
\end{theorem}

\section{Numerical Experiments}\label{sect:exp}
We test the proposed algorithm on both simulated data and real EEG/MEG data to illustrate its effectiveness. The simulation study is based on a real head model. Although we know noise exists in both electrodes and brain source space, to what extent different levels of noise impacts on the inverse solvers are unclear or less studied so far. We consider various levels of signal-to-noise ratio (SNR) in both sensor and source spaces to validate the proposed algorithm in comparison with some popular algorithms in   the ESI literature \cite{hamalainen1994interpreting}\cite{uutela1999visualization}\cite{pascual2002standardized}\cite{gramfort2012mixed}. The SNR is defined by $\text{SNR} = 10\log_{10}({P_s}/{P_n})$,
where $P_s$ is the power of signal and $P_n$ is the power of noise signal. 

\subsection{Competing algorithms and error metrics}
We compare our method with four benchmark algorithms, namely MNE \cite{hamalainen1994interpreting}, MCE~\cite{uutela1999visualization}, sLORETA~\cite{pascual2002standardized}, and MxNE~\cite{gramfort2012mixed}. For the MCE solver, we employ the Homotopy algorithm due to its superior performance  observed in our previous work on ESI \cite{liu2017graphTBD}. In addition, we find that incorporating patial smoothness with the $\ell_1$ approach in MCE improves the reconstruction results, so we adopt this improved version of MCE in experiments. As for 
MxNE \cite{gramfort2012mixed}, we impose an $\ell_1$ norm on the source space and an $\ell_2$ across time, which is one form of MxNE, thus leading to $\ell_{21}$ regularization with an assumption of fixed source orientations and spatial smoothness. In this paper, we use $\ell_{21}$ instead of MxNE for a precise description. We find that Algorithm \ref{alg} is insensitive to parameters, so in our experiments, we fix
$\lambda=3$, $\beta=3$, $\alpha=0.01$, $\gamma_1=\gamma_2$ and tune $K \in \{ 60, 120\}$ and $\gamma_1 \in \{0.01, 0.001\}$. We set the sparsity penalty parameter for MCE to be $0.01$ and $\ell_{21}$ is set to be $0.05$ respectively.

We quantitatively evaluate the performance of each competing algorithm based on  the following  metrics:
\begin{itemize}
	\item Data fitting (DF). We define 
	\begin{align*}
	E_{tot} = \sum_{i=1}\norm{\mathbf x_i-\bar{\mathbf x}}_2^2 \quad \mbox{and} \quad
	E_{res} = \sum_{i=1}\norm{\mathbf x_i-\hat{\mathbf x}}_2^2,
	\end{align*}
	where $\mathbf x_i$ is $i$-th column in the EEG data  $X$,  $\bar{\mathbf x}$ is the mean of $X$ along the time axis, and $\hat{\mathbf x}_i$ is the fitted value defined as $\hat{\mathbf x}_i =L\hat{\mathbf s}_i$ for any reconstructed source signal at time $i$.  {The DF metric} is defined as $r^2 = |1 - \frac{E_{res}}{E_{tot}}|$.
	\item Reconstructed error (RE) in source location is defined as 
	$\mbox{RE}=\frac{\norm{\hat S -S}_F^2}{\norm{S}_F^2}$. However, RE cannot completely characterize the reconstruction performance due to the ESI's inherent difficulty of having a highly correlated lead field matrix. In particular, one often observes that a reconstructed source is located in the neighborhood of the true location, thus leading to a very large RE. 
	\item Localization error (LE) measures the geodesic distance between two source locations on the cortex meshes using the Dijkstra shortest path algorithm.
	% \cite{dijkstra1959note}. 
	%The reconstructed source location is selected as the  one with the largest magnitude. 
	As two hemispheres of a brain are disconnected, we calculate LE separately for each hemisphere. 
	\item Area under curve (AUC) \cite{molins2008quantification} is particularly useful to characterize the overlap of an extended source activation pattern\cite{zhu2014reconstructing}\cite{sohrabpour2016imaging}.
\end{itemize}
For the four metrics, better reconstruction results are expected if DF, AUC are closer to 1 and RE, LE are closer to 0.

\vspace{-1em}
\subsection{Simulation study based on a real head model}
As in many brain imaging problems, we do not know the underlying ground truth. As a result, we rely on 
synthetic data by numerical simulations to characterize the performance of various algorithms. In this paper, we consider a real head model with simulated brain source signals. The head model was reconstructed from T1-MRI images of a male subject scanned at 26 years old in Massachusetts General Hospital. We used a 128-channel BioSemi EEG cap layout and co-registered it with the subject's head surfaces (Electrodes layout and coregistration can be found in Appendix Fig.\ref{coregistration}). Brain structure segmentation and cortical surface reconstruction were conducted using FreeSurfer. Coregistration of the head surface and EEG electrodes were conducted using Brainstorm~\cite{tadel2011brainstorm} and then verified using the GUI of coregistration function in MNE-Python \cite{gramfort2014mne}.
%MNE-Python was used to build the forward model. The minimum distance between sources and the inner skull surface was set to 5 mm. 
The conductivity of brain, skull, and scalp were set to 0.3 S/m, 0.006 S/m and 0.3 S/m, respectively. 
%The octahedral subdivision grade was set to 5. 
The source space contained 1026 sources in each hemisphere, with 2052 sources in total.

In the simulation, we randomly selected two activation locations, one from each hemisphere so that LE could be calculated without confusion. We then generated two active source signals via a 5th-order autoregressive (AR) model at these locations (see Fig.\ref{fig:brain_activation}). In the simulated experiments, we considered three states with different activation locations, with the sampling frequency set to 100 Hz, and the time window set to 2 s. A detailed description of the simulated signals can be found in Ref.\cite{haufe2016simulation}. Each of the three states corresponds to a multi-channel time series of length 200.  As a result, we obtain a time series with a length of 600 in total. We consider additive white noise with various SNR levels in both channel and source spaces, denoted as  $\mbox{SNR}_{C}$ and  $\mbox{SNR}_{S}$, respectively. We examine three noise levels in each space: $\mbox{SNR}_{C}= \mbox{30 dB, 20 dB, and } 10$ dB and $\mbox{SNR}_{S} = \infty \mbox{ (noiseless)}, 30$ dB, and 10 dB. The impact of noises on the noiseless signal can be found in Fig.\ref{fig:noise_impact} and  Fig.\ref{fig:noise_impact_part2}.

\begin{figure*}
	\centering
	\begin{subfigure}[b]{0.31\textwidth}
		\centering
		\includegraphics[width=1\textwidth,height=0.65\textwidth]{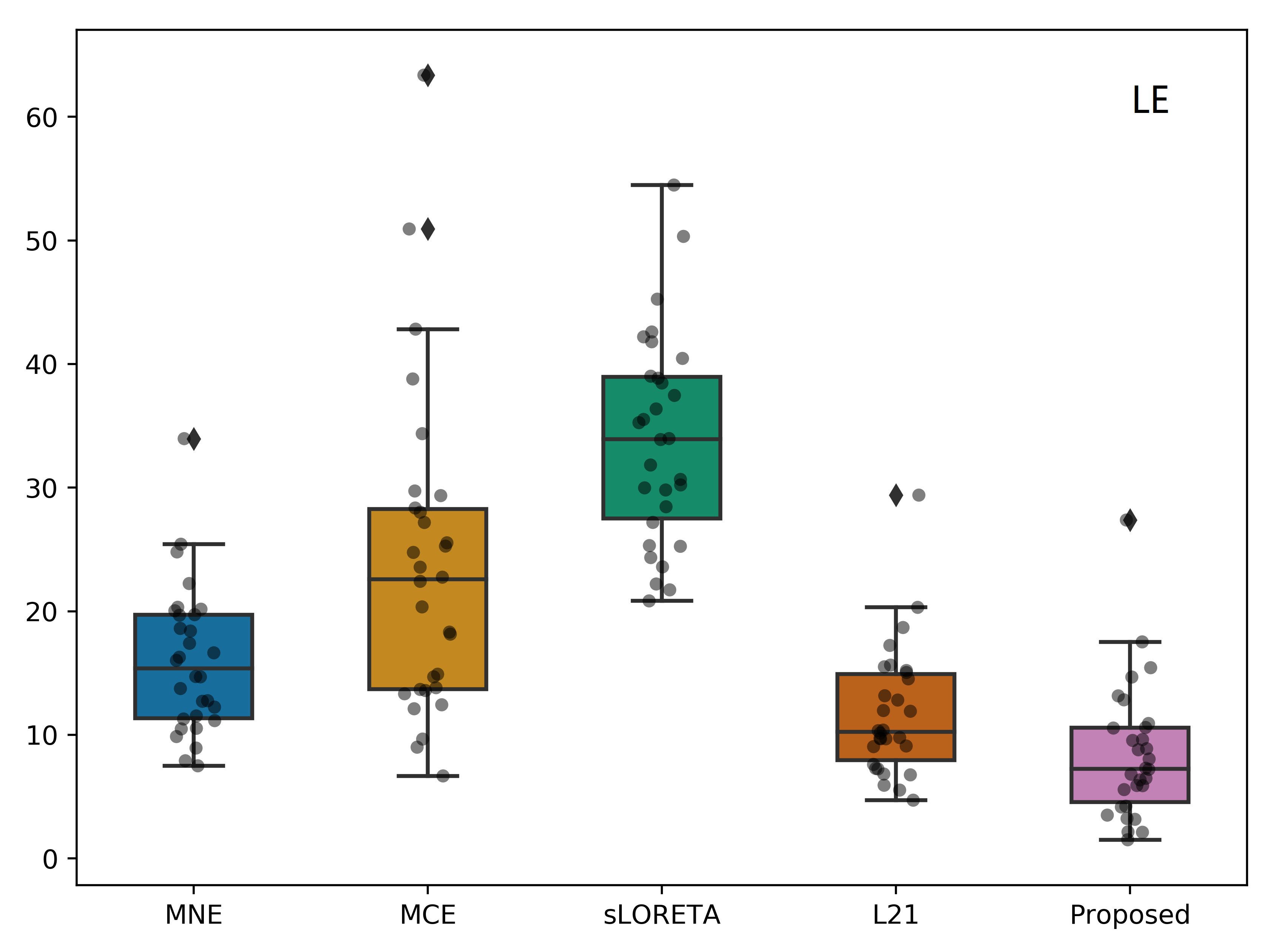}
		%	\caption[] {}
		%  \caption[] {{\small $L_1$ under $D_1$}} 
		%		\label{fig:f1_1d1s_etas}
	\end{subfigure}
	\hspace{0em}
	\begin{subfigure}[b]{0.31 \textwidth}  
		\centering 
		\includegraphics[width=1\textwidth,height=0.65\textwidth]{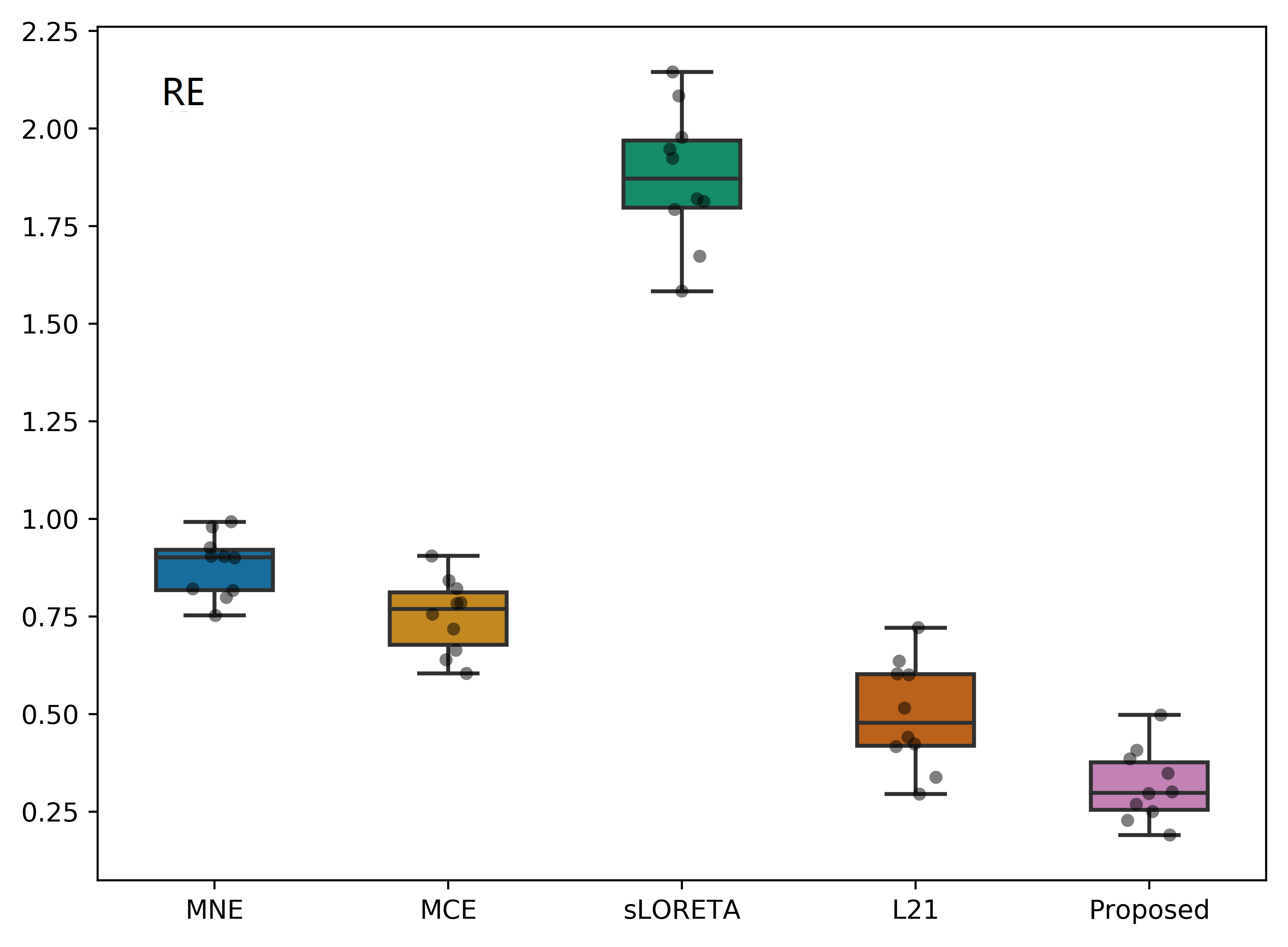}
		%	\caption[] {}
		%  \caption[] {{\small $L_1$ under $D_2$}}    
		%	\label{fig:f1_4d1s_etas}
	\end{subfigure}
	% \hspace{0em}
	%	\vspace{0em}
	\begin{subfigure}[b]{0.31 \textwidth}  
		\centering 
		\includegraphics[width=1\textwidth,height=0.65\textwidth]{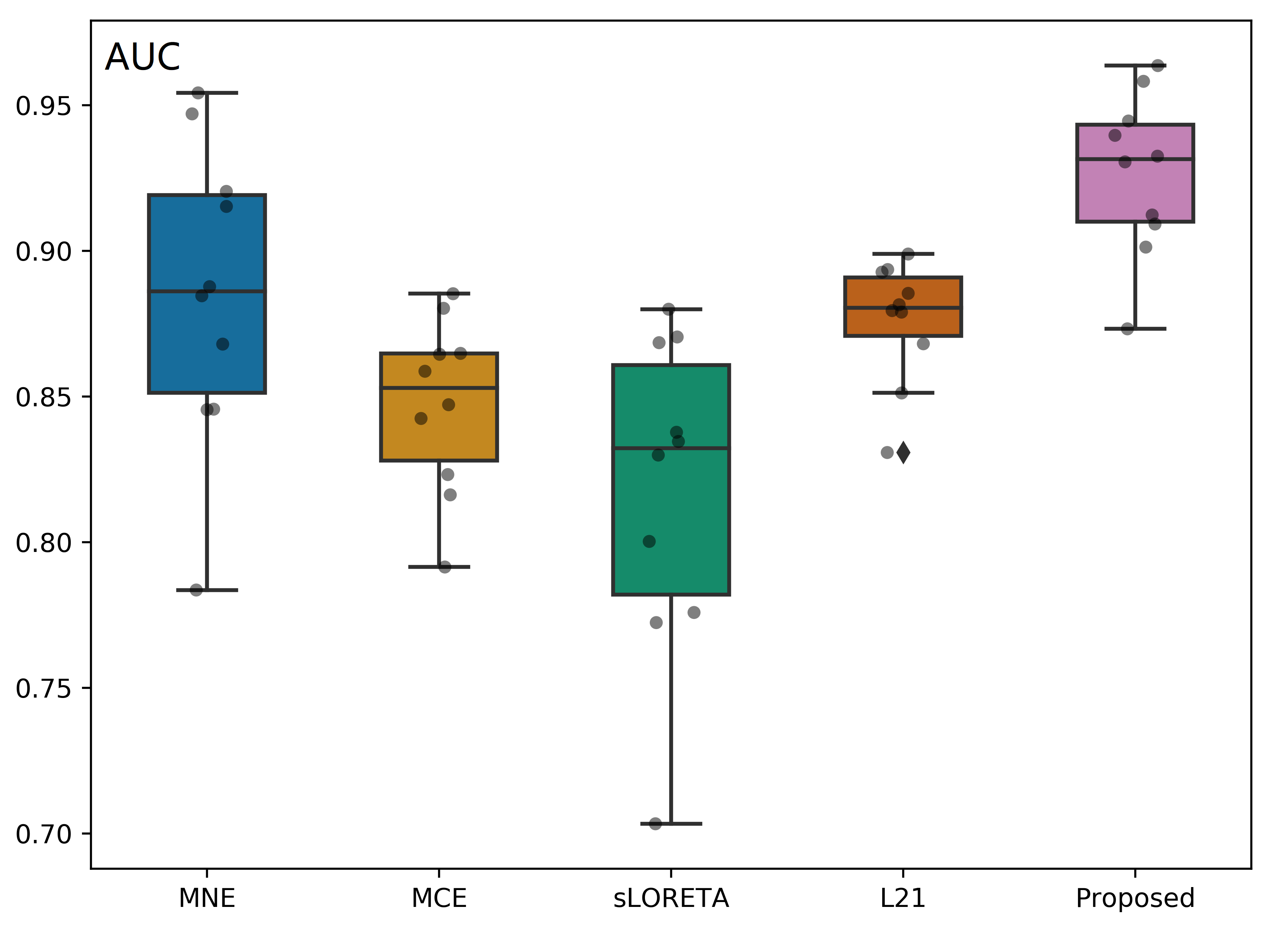}
		%	\caption[] {}
		%  \caption[] {{\small $L_1$ under $D_3$}}    
		%	\label{fig:f1_6d1s_etas}
	\end{subfigure}
	\vspace{0em}
	\begin{subfigure}[b]{0.31\textwidth}
		\centering
		\includegraphics[width=1\textwidth,height=0.65\textwidth]{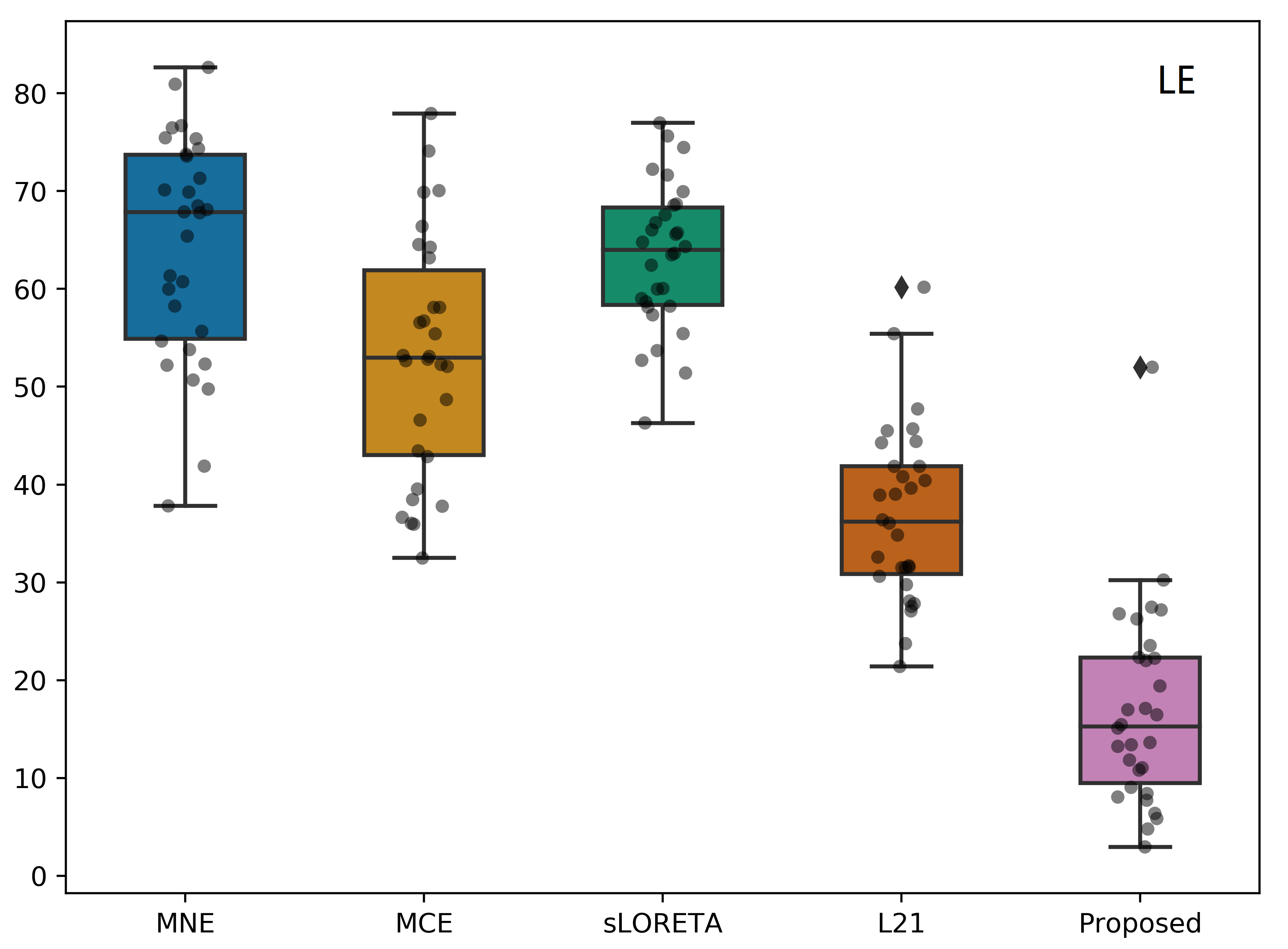}
		%	\caption[] {}
		%  \caption[] {{\small $L_1$ under $D_1$}} 
		%		\label{fig:f1_1d1s_etas}
	\end{subfigure}
	\hspace{0em}
	\begin{subfigure}[b]{0.31 \textwidth}  
		\centering 
		\includegraphics[width=1\textwidth,height=0.65\textwidth]{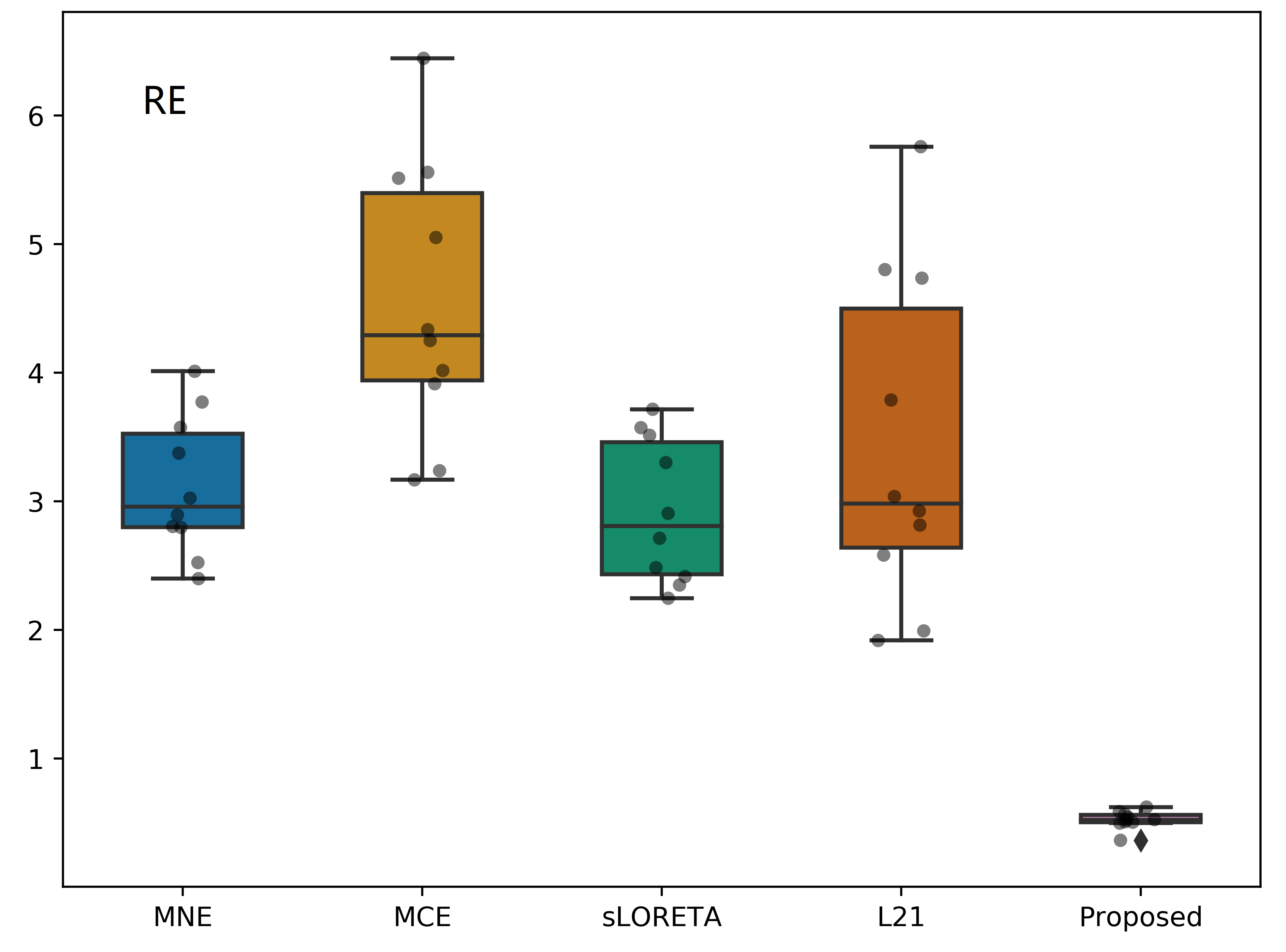}
		%	\caption[] {}
		%  \caption[] {{\small $L_1$ under $D_2$}}    
		%	\label{fig:f1_4d1s_etas}
	\end{subfigure}
	\hspace{0em}
	%	\vspace{0em}
	\begin{subfigure}[b]{0.31 \textwidth}  
		\centering 
		\includegraphics[width=1\textwidth,height=0.65\textwidth]{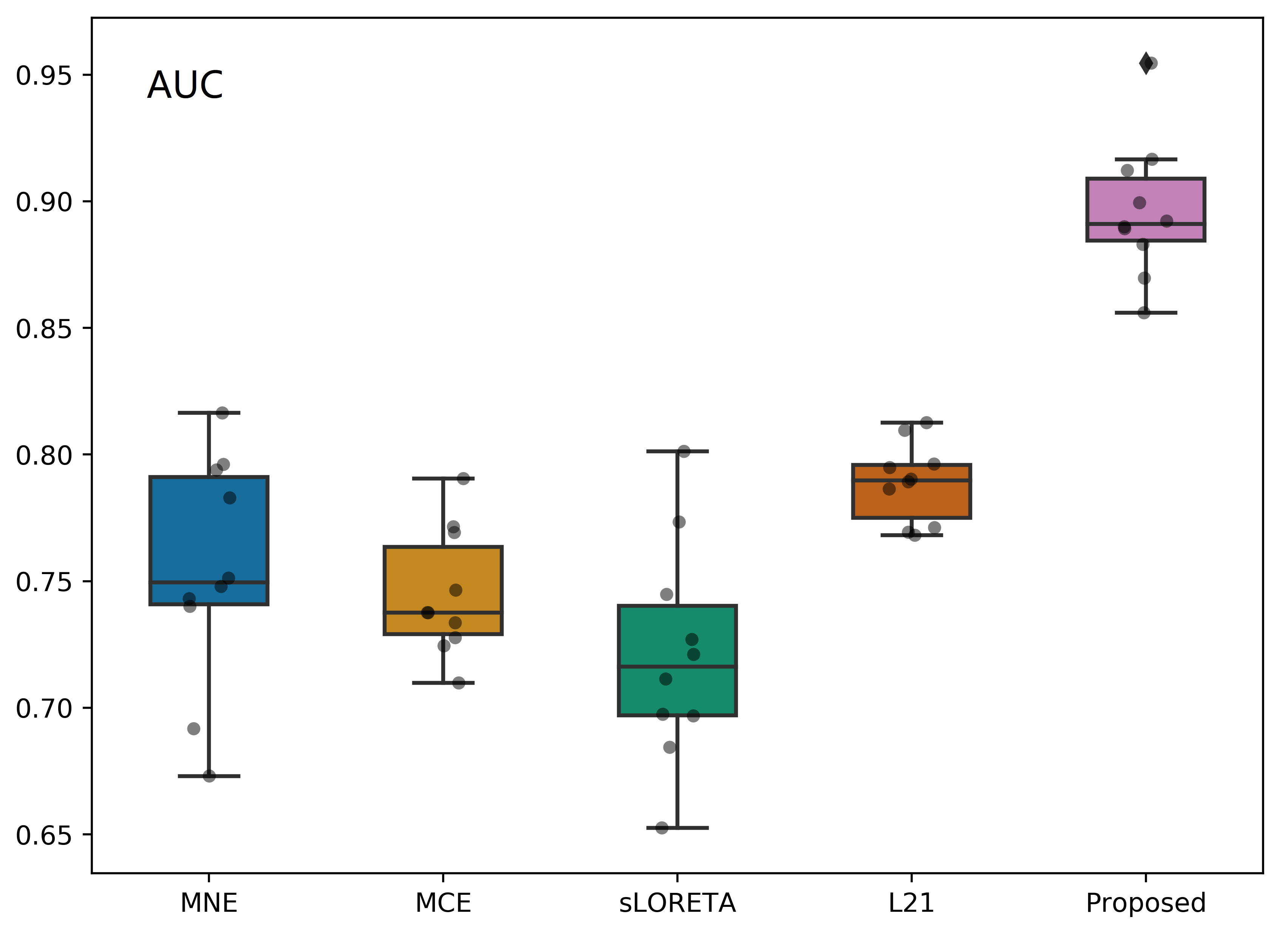}
		%	\caption[] {}
		%  \caption[] {{\small $L_1$ under $D_3$}}    
		%	\label{fig:f1_6d1s_etas}
	\end{subfigure}
	\caption{\small LE (left), RE (middle),  and AUC (right) comparison. Top: $\mbox{SNR}_{C}$ = 20 dB/$\mbox{SNR}_{S}$ = 10 dB. Bottom:  $\mbox{SNR}_{C}$ = 10 dB/$\mbox{SNR}_{S}$ = 10 dB.}
	\label{fig:boxplot_comparison}
	\vspace{-1em}
\end{figure*}

The results of the proposed algorithm in comparison to the benchmark ones are summarized in Tables~\ref{table:result_t_snr30}-\ref{table:result_t_snr10} according to different sensor noise levels. Each reported value is the mean result from 10 repeated experiments. The results show that the proposed approach outperforms the benchmark algorithms in most cases, particularly excelling in terms of LE and AUC.
Specifically in Table~\ref{table:result_t_snr30} for a higher $\mbox{SNR}_{C}$ value (less noise in the channel space), all the algorithms perform very well in terms of LE. The proposed algorithm is marginally better than sLORETA (2.92 mm versus 3.64 mm). Although MCE has the best performance on DF, it has the largest LE compared to other algorithms. This phenomenon suggests that small DF does not necessarily lead to good performance in terms of RE, LE and AUC. 
In Table~\ref{table:result_t_snr20}, we notice that  sLERETA, $\ell_{21}$ and our algorithm are robust to noise, while MCE and MNE have a pronounced deterioration in performance, when $\mbox{SNR}_{C}$ is decreased from 30 dB to 20 dB. Table~\ref{table:result_t_snr10} shows significant improvements of the proposed approach over the state-of-the-art in terms of LE and AUC. The mean LE of our algorithm is between 13 mm and 15 mm, compared favorably with 30-37 mm for sLERETA, 58-68 mm for $\ell_{21}$, and 45-65 mm for MNE and MCE. The AUC of the proposed algorithm is about 0.9, while the AUC values of other algorithms are all below 0.8. 
Finally, we illustrate the performance of LE, RE and AUC using boxplots in Fig.~\ref{fig:boxplot_comparison} for two cases when $\mbox{SNR}_{C}$ = 20 dB, $\mbox{SNR}_{S}$ = 10 dB (top panel) and  $\mbox{SNR}_{C}$ = 10 dB, $\mbox{SNR}_{S}$ = 10 dB (bottom panel). The boxplots demonstrate that the proposed algorithm achieves the best results with the least variance among different random realizations.

\begin{table*}[h] 	
	\caption{Performance Summary for $\mbox{SNR}_{C}=30$ dB} 
	\centering
	%	\begin{scriptsize}
	\scalebox{0.75}{
		\begin{tabular}{c|cccc|cccc|cccc } \hline \hline
			\multirow{2}{*}{Methods} & \multicolumn{4}{c|}{{ $\mbox{SNR}_{S}$= inf dB} } & \multicolumn{4}{c|}{{ $\mbox{SNR}_{S}$ = 30 dB} }   &  \multicolumn{4}{c}{{ $\mbox{SNR}_{S}$ = 10 dB}}   \\ \cline{2-13}		
			& {   DF}& {   RE} & {   LE}  & {   AUC} & {   DF } & {   RE }& {   LE }   & {   AUC}   & {   DF } & {   RE }& {   LE }   & {   AUC}   \\ 	\hline 
			MNE	          & 0.952     &  0.62        &  8.72 & 0.927                & 0.943  & 0.67           & 9.70  &  0.920                       & 0.901  &  0.70  &  16.13 & 0.887  \\  
			MCE           & \textbf{0.983}   &  0.21 &  12.86  & 0.902              & \textbf{0.982}  & 0.21  &  12.82   &  0.910                    & 0.833  & 4.46  &   17.90  & 0.720   \\   
			sLORETA       &  0.418 & 1.44  &   3.64  & 0.936                        & 0.401  & 1.40           &  4.39  & 0.922                       & 0.396  &  1.66  & {8.26} & 0.881     \\  
			$\ell_{21}$           &  0.972 &  0.33 &  5.77 & 0.965                          & 0.972  & 0.30           &  5.52   &  0.960                     &\textbf{ 0.928}  &  0.44  &  10.88 & 0.905   \\  
			proposed      &  0.969 &  \textbf{0.15} & \textbf{2.92}  &\textbf{ 0.981}        & 0.968  &\textbf{ 0.16 } & \textbf{ 3.34 }  &\textbf{ 0.980 }    & 0.903  &  \textbf{0.30}  & \textbf{8.25} &\textbf{ 0.911}   \\
			\hline 			
		\end{tabular}
	}
	%	\end{scriptsize}
	\label{table:result_t_snr30}
\end{table*}

\begin{table*}[h] 	
	\caption{Performance Summary for  $\mbox{SNR}_{C}$ = 20 dB} 
	\centering
	\scalebox{0.75}{
		\begin{tabular}{c|cccc|cccc|cccc } \hline \hline
			\multirow{2}{*}{Methods} & \multicolumn{4}{c|}{{   $\mbox{SNR}_{S}$ = inf dB} } & \multicolumn{4}{c|}{{   $\mbox{SNR}_{S}$ = 30 dB} }   &  \multicolumn{4}{c}{{   $\mbox{SNR}_{S}$ = 10 dB}}   \\ \cline{2-13}		
			& {   DF}& {   RE} & {   LE}  & {   AUC} & {   DF } & {   RE }& {   LE }   & {   AUC}   & {   DF } & {   RE }& {   LE }   & {   AUC}   \\ 	\hline
			MNE	    & 0.901  &  0.85  & 16.45  & 0.854                             &   0.902  & 0.85  & 18.28  & 0.862                               & 0.901  & 0.88  &  22.59  & 0.847     \\  
			MCE  & \textbf{0.945}  &  0.71  & 29.23  & 0.826                    &  \textbf{0.945}  & 0.77  & 32.46   &  0.829               & \textbf{0.943}  & 0.75  &  33.93  & 0.817     \\   
			sLORETA   & 0.395  &  1.55  & 8.71 & 0.891                               &   0.406  & 1.67  & 10.01   & 0.896                              & 0.396  & 1.88  &  10.24 & 0.876     \\  
			$\ell_{21}$   & 0.926  &  0.41  & 9.75 & 0.913                       &   0.927  & 0.42  & 10.49  & 0.924                                & 0.928  & 0.50  &  15.37  & 0.885     \\  
			proposed   & 0.908  &\textbf{0.31} & \textbf{5.14} & \textbf{0.958}       &   0.909  &\textbf{0.326}  & \textbf{4.95} &\textbf{0.966}           & 0.903  & \textbf{0.32}  & \textbf{7.25}  & \textbf{0.927}     \\
			\hline 			
		\end{tabular}
	}
	%	\end{scriptsize}
	\label{table:result_t_snr20}
\end{table*}

\begin{table*}[h] 	
	\caption{Performance Summary for $\mbox{SNR}_{C}$ = 10 dB} 
	\centering
	%	\begin{scriptsize}
	\scalebox{0.75}{
		\begin{tabular}{c|cccc|cccc|cccc } \hline \hline
			\multirow{2}{*}{Methods} & \multicolumn{4}{c|}{{   $\mbox{SNR}_{S}$ = inf dB} } & \multicolumn{4}{c|}{{   $\mbox{SNR}_{S}$ = 30 dB} }   &  \multicolumn{4}{c }{{   $\mbox{SNR}_{S}$ = 10 dB}}   \\ \cline{2-13}		
			& {   DF}& {   RE} & {   LE}  & {   AUC} & {   DF } & {   RE }& {   LE }   & {   AUC}   & {   DF } & {   RE }& {   LE }   & {   AUC}   \\ 	\hline
			MNE	            & 0.741  & 2.80  & 50.87  & 0.753                    &  0.736 &  2.93 & 46.35  & 0.751                                 & 0.738  &  3.12  & 52.95 & 0.745 \\  
			MCE     &\textbf{ 0.833}  & 4.46  & 64.97  & 0.720                  & \textbf{0.833 } &  4.58 & 60.63  & 0.733               & \textbf{0.831}  &  4.55    & 63.98  & 0.721 \\   
			sLORETA         & 0.416  & 2.49  & 33.69 & 0.795                    & 0.395  &  2.63 & 31.49  & 0.785                                 &  0.401 & 2.92      & 36.22  & 0.789\\  
			$\ell_{21}$             & 0.825  & 2.89  & 62.37  & 0.766                    & 0.825  &  2.91 & 58.83  & 0.771                                  & 0.828  & 3.43      & 67.82  & 0.754 \\  
			proposed   & 0.736  & \textbf{0.53}  & \textbf{14.64}  & \textbf{0.908}  & 0.736  & \textbf{ 0.52} & \textbf{13.65} & \textbf{0.898}  & 0.734  & \textbf{ 0.52}    & \textbf{ 15.28} & \textbf{0.896}\\
			\hline 			
		\end{tabular}
	}
	%	\end{scriptsize}
	\label{table:result_t_snr10}
\end{table*}

%reconstructed activation pattern from different algorithms (from left to right): proposed, MCE, $\ell_{21}$, MNE, sLORETA. the activation is in the .  
%Lower row: $\mbox{SNR}_{C}$ = 10 dB and $\mbox{SNR}_{S}$ = 10 dB, the ground true source activation (left most) and reconstructed activation pattern from different algorithms (from left to right): proposed, MCE, $\ell_{21}$, MNE, sLORETA, the activation is in the lateral occipital area.

\begin{figure}[h]
	\centering
	\begin{subfigure}[b]{0.8\textwidth} 
		\includegraphics[width=\linewidth,height=0.15\linewidth]{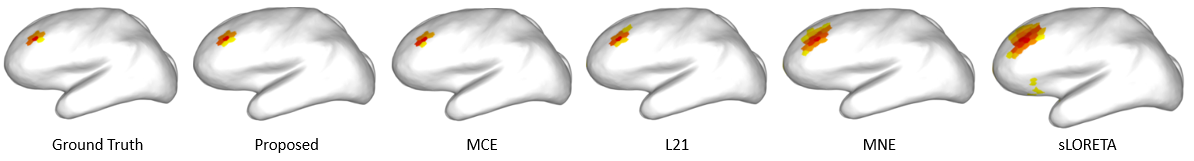}%
	\end{subfigure}
	\begin{subfigure}[b]{0.8\textwidth} 
		\includegraphics[width=\linewidth,height=0.2\linewidth]{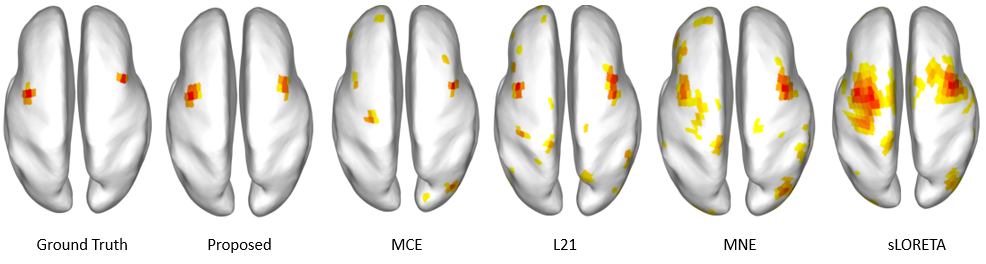}%
	\end{subfigure}
	%	\vspace{0em}
	\begin{subfigure}[b]{0.8\textwidth} 
		\includegraphics[width=\linewidth,height=0.16\linewidth]{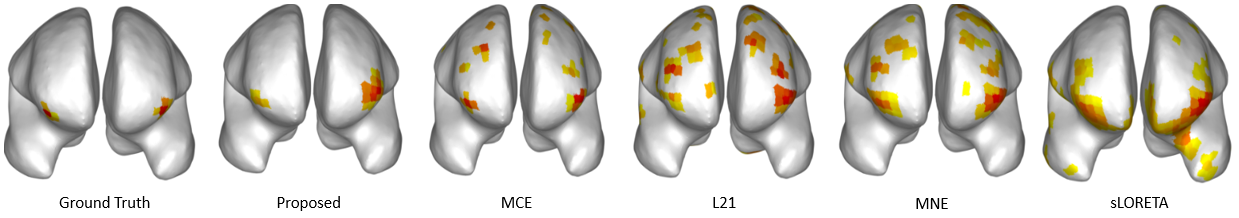}%
	\end{subfigure}
	\caption[]{Source reconstruction comparison with different noise levels: $\mbox{SNR}_{C}$ = 20 dB and $\mbox{SNR}_{S}$ = 30 dB (top),
		% the activation is in the Rostral middle frontal area.
		$\mbox{SNR}_{C}$ = 10 dB and $\mbox{SNR}_{S}$ = 30 dB (middle), and 
		%the activation is in the caudal middle frontal area.  
		$\mbox{SNR}_{C}$ = 10 dB and $\mbox{SNR}_{S}$ = 10 dB (bottom). 
		% and the activation is in the lateral occipital area. For all 3 rows, 
		%		 The ground true source activation (left most) and reconstructed activation pattern from different algorithms (from left to right): our proposed algorithm, MCE, $\ell_{21}$, MNE, sLORETA.
	}
	\label{fig:cortex_comparison}
	%\vspace{-1em}
\end{figure}

We illustrate the results of source reconstruction under different noise levels and locations in Fig.~\ref{fig:cortex_comparison}. The top row of Fig.~\ref{fig:cortex_comparison} is for $\mbox{SNR}_{C}$ = 20 dB/$\mbox{SNR}_{S}$ = 30 dB  with  true activation  in the rostral middle frontal area. In this case of high SNR, all the methods achieve satisfactory results, except for over-diffused solutions from the MNE and sLORETA reconstructions. We then increase  noise  to $\mbox{SNR}_{C}$ = 10 dB/$\mbox{SNR}_{S}$ = 30 dB and generate the ground true source activation in caudal middle frontal area. The reconstruction results are given in the middle row of Fig.~\ref{fig:cortex_comparison}, which shows that the activation patterns from the proposed algorithm is the closest to the ground-truth. MCE and $\ell_{21}$  can correctly identify some locations, but with a lot of spurious activations on the cortex surface. On the other hand, MNE and sLORETA give the over-diffused solution, but the locations with largest magnitude on both hemisphere align well with the true activation locations. The bottom row in Fig.~\ref{fig:cortex_comparison} is the case of larger noise, namely $\mbox{SNR}_{C}$ = 10 dB and $\mbox{SNR}_{S}$ = 10 dB, where the proposed algorithm yields a sparse and accurate reconstruction.
 
\par In summary, numerical results on synthetic EEG data show that the proposed approach is comparable to the benchmark algorithms at lower noise levels. When the noise level is high, we demonstrate significant improvements over the benchmark algorithms. The observation confirms that source signal denoising and graph structure play an important role in identifying activation patterns under a high level of noise in both sensor and source spaces.

\vspace{-1em}
\subsection{Real data experiment}
We conduct experiments on a real dataset that is publicly accessible through the MNE-Python package~\cite{gramfort2014mne}. The EEG/MEG data is collected when the subjects are given auditory and visual stimuli. In the experiment, the subject is presented with checkerboard patterns to the left or right eye, and interspersed by high frequency noise to the left or the right ear. 
The recording device is a whole-head Elekta Neuromag Vector View 306 MEG system with 102 triple-sensor elements (two orthogonal planar gradiometers and one magnetometer per location). The EEG data from a 60-channel cap are also recorded simultaneously. Standard preprocessing steps including bandpass filter, bad epoch rejection, and bad channel selections were conducted before applying ESI algorithms\cite{gramfort2013time}. 
There are 7498 sources distributed over the cortical surfaces.

\begin{figure}[h]
	\centering
	\begin{subfigure}[b]{0.75\textwidth} 
		\includegraphics[width=\linewidth,height=0.27\linewidth]{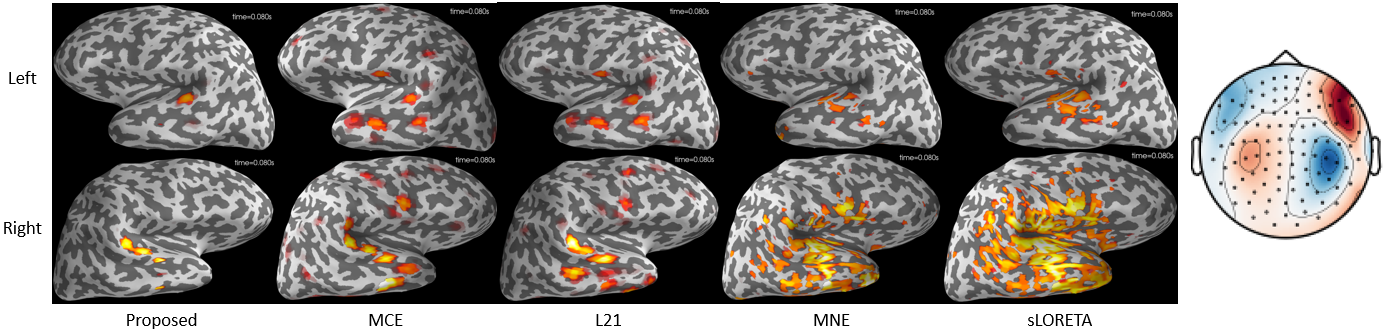}%
	\end{subfigure}
	\begin{subfigure}[b]{0.75\textwidth} 
		\includegraphics[width=\linewidth,height=0.27\linewidth]{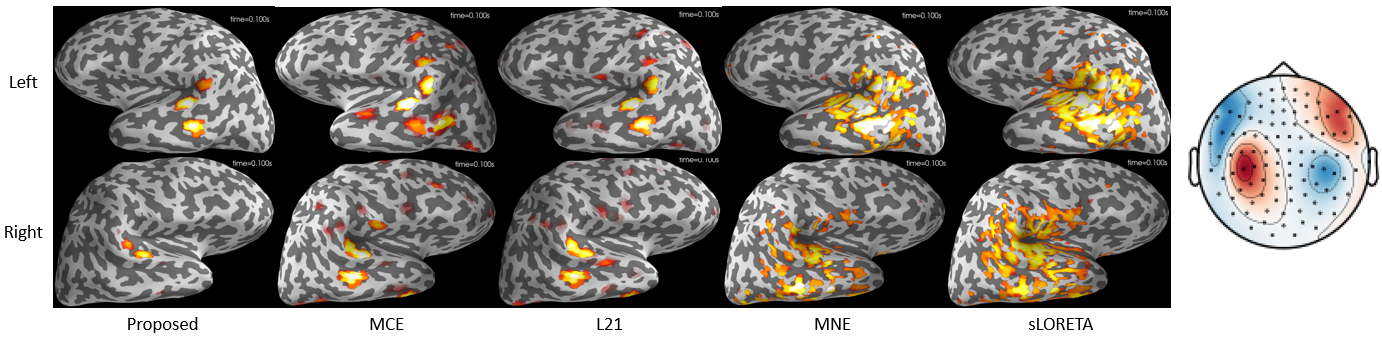}%
	\end{subfigure}
	\begin{subfigure}[b]{0.75\textwidth} 
		\includegraphics[width=\linewidth,height=0.27\linewidth]{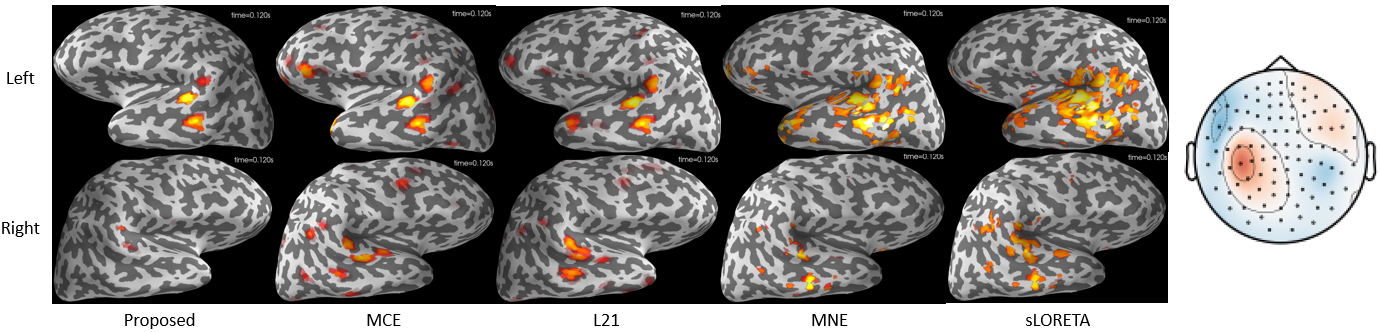}%
	\end{subfigure}
	\caption[]{Source activation patterns from MEG data at $t~=~0.08$~s (top), $t~=~0.1$~s (middle), $t~=~0.12$~s (bottom).}
	\label{fig:source_readdata_meg}
	\vspace{-1em}
\end{figure}

The epochs under study are from left auditory stimuli (LAS) and left visual stimuli (LVS). There are 66 epoches for LAS and 73 for LVS. The time used for all the epoches of LVS and LAS is from 0.03 s to 0.16 s after the stimuli events. The average time course aligned with stimuli events for EEG, Gradiometers and Gagnetometers are illustrated in Fig.~\ref{fig:tsplots} for LAS and Fig.\ref{fig:tsplots_lv} for the LVS in Appendix. By checking the time series plots in Fig.~\ref{fig:tsplots}, the event related potential (ERP) activation pattern for LAS is very clear from time 0.08 s to 0.12 s and the related topomaps are illustrated on rightmost plot in Fig.\ref{fig:source_readdata_meg} for the time point 0.08 s, 0.10 s and 0.12 s. 
Following the same parameter setting as the simulation study, we perform the source localization on an averaged epoch in a moving window with 15 epoches (overlap=5) for LAS and LVS and the final activation patterns on the cortex for LAS are illustrated in Fig.~\ref{fig:source_readdata_meg}
presents the source reconstruction results for  $t=0.08$~s (top), $t=0.1$ s (middle), and $t=0.12$~s (bottom), respectively. At time $t=0.08$~s, all the algorithms show that the right hemisphere has stronger activated sources than the left hemisphere, which is consistent with the topomaps in Fig.~\ref{fig:source_readdata_meg} where the right hemisphere has a stronger electricity potential than the left hemisphere. At time $t=0.1$~s the left hemisphere demonstrates a stronger activation. At time $t=0.12$ s, we observe that the activation in the left hemisphere is stronger than the right hemisphere from both reconstructed source space and the topomaps in sensor space. The other competing methods suffer from  over-diffused or spurious activations in their reconstruction, similar to our observation for the synthetic data.

\begin{figure}[H]
	\centering
	\includegraphics[width=\linewidth,height=0.4\linewidth]{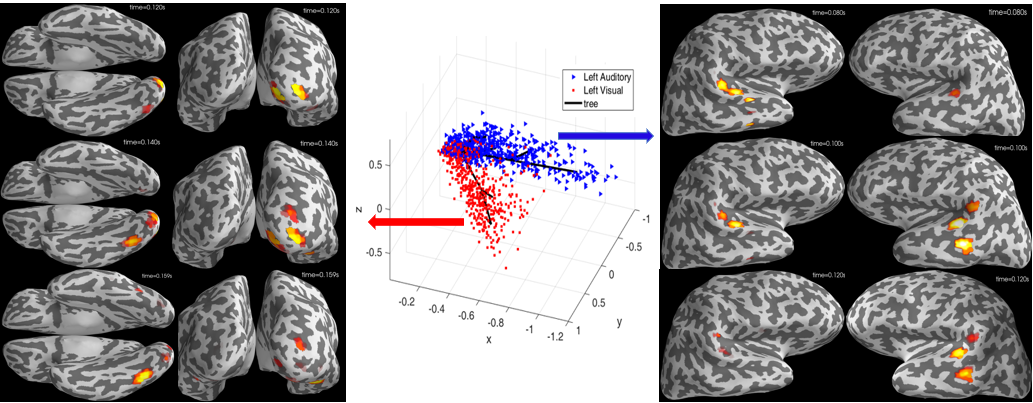}%
	\caption[]{Reconstructed sources and spanning tree (middle figure) from landmarks projected to the first 3 principal components for LAS (in blue) and LVS (in red), and the corresponding cortex activation for different time points for LAS (left) and LVS (right).}
	\label{fig:landmarks}
\end{figure}
\vspace{-1em}

To visualize and validate the existence of spanning tree in real data, we project the source solutions of all the averaged epoches from different moving windows for LAS and LVS on the first 3 principal components in Fig.~\ref{fig:landmarks}. We can see the intersection of scattered points from LAS and LVS denotes the patterns before the actual ERP activations and then the activations under different stimuli condition progress on different branches of the spanning tree. 

Compared to all the benchmark algorithms, our approach provides a sparse source reconstruction with less spuriously activated sources incurred from the noise and activation patterns and the activations are more consistent across time.

\section{Conclusions} \label{sect:conclude}

In this paper, we presented a novel probabilistic ESI model. Particularly important is the introduction of a hierarchical graph prior that blends the flexibility of maintaining consistency across time as well as allowing the existence of distinct representative landmark activation patterns in the studied period. An efficient algorithm based on alternating convex search was proposed 
with provable convergence.  
We conducted extensive numerical experiments, including both the synthetic data and real data, and demonstrated that the proposed algorithm can robustly localize the activated sources with satisfactory precision to handle large noise in both channel and source spaces, where the traditional algorithms often fail to do so. Specifically for synthetic experiments, our algorithm outperforms all the benchmark algorithms and yields significant improvements when the noise level in channel space and source space is high. This study also filled an gap in the literature to investigate the impact of noise separately from channel and source spaces. In the example of real data, our algorithm rendered  more consistent   reconstructions,   while allowing variations across the time axis. Compared with other algorithms, the source signal reconstructed by ours is  less contaminated from spurious noise originated from brain spontaneous source activations or channel noise.

\bibliographystyle{unsrt}
\bibliography{EEG-graph}

\clearpage

\onecolumn
%\appendices
\section{Appendix}
\setcounter{figure}{0}
\counterwithin{figure}{section}
\begin{figure*}[h]
	\centering
	\begin{minipage}{.27\textwidth}
		\centering
		\includegraphics[width=\linewidth]{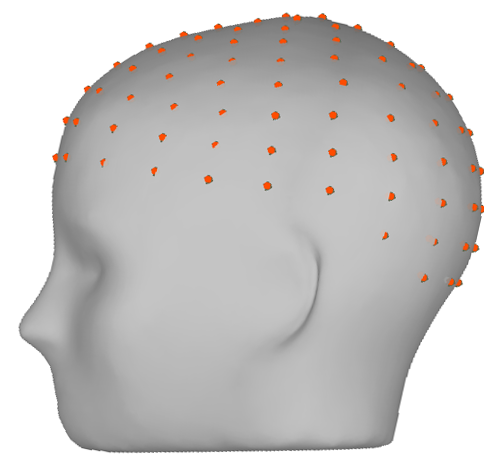}
		%\caption{}
		%    \caption{Summary of reconstruction error for different $\lambda$ and $\beta$}
		\label{Over_all_result}
	\end{minipage}%
	\hspace{2em}
	\begin{minipage}{.27\textwidth}
		\centering
		\includegraphics[width=\linewidth]{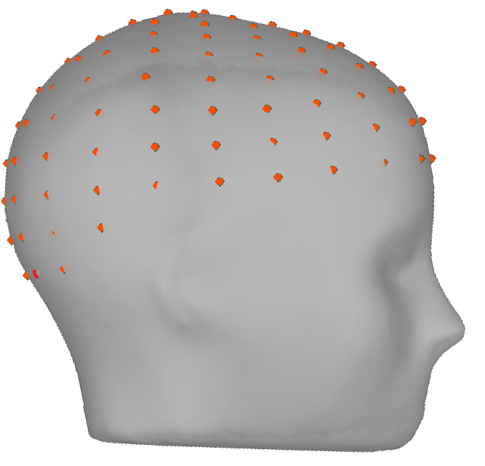}
		%\caption{}
		%    \caption{Rank summary}
		\label{rank_summary}
	\end{minipage}
	\begin{minipage}{.27\textwidth}
		\centering
		\includegraphics[width=0.8\linewidth, height=0.9\linewidth]{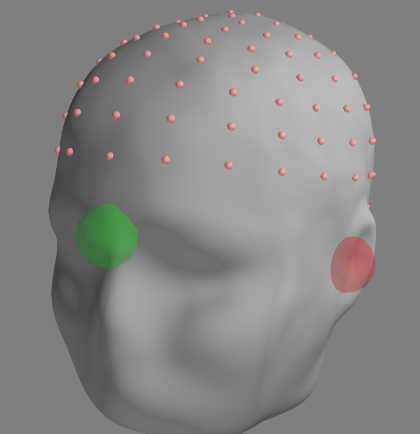}
		%\caption{}
		%    \caption{Rank summary}
		%	\label{rank_summary}
	\end{minipage}
	\caption{EEG channel layout of BioSemi 128 system (left two figures), and coregistration with head surfaces in MNE-Python (figure on the right)}
	\label{coregistration}
\end{figure*}

\begin{figure*}[h]
	\centering
	\includegraphics[clip,width=0.65\columnwidth]{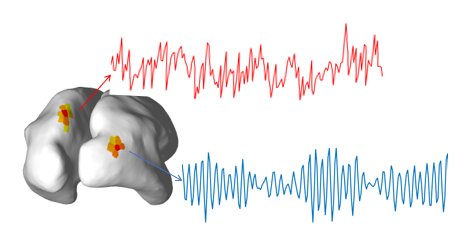}%
	\caption{Activated brain source extents with autoregressive time courses.}
	\label{fig:brain_activation}
\end{figure*}

\begin{figure*}[h]	
	\centering	
	\begin{subfigure}[b]{0.8\textwidth} 
		\centering
		\includegraphics[width=0.75\textwidth,height=0.18\textheight]{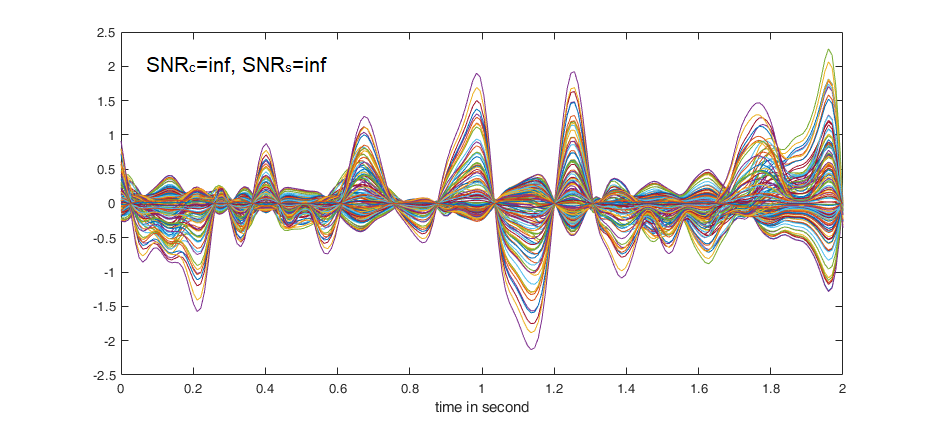}%
	\end{subfigure}
	%	\vspace{5em}		
	\begin{subfigure}[b]{0.8\textwidth} 
		\centering
		\includegraphics[width=0.75\textwidth,height=0.18\textheight]{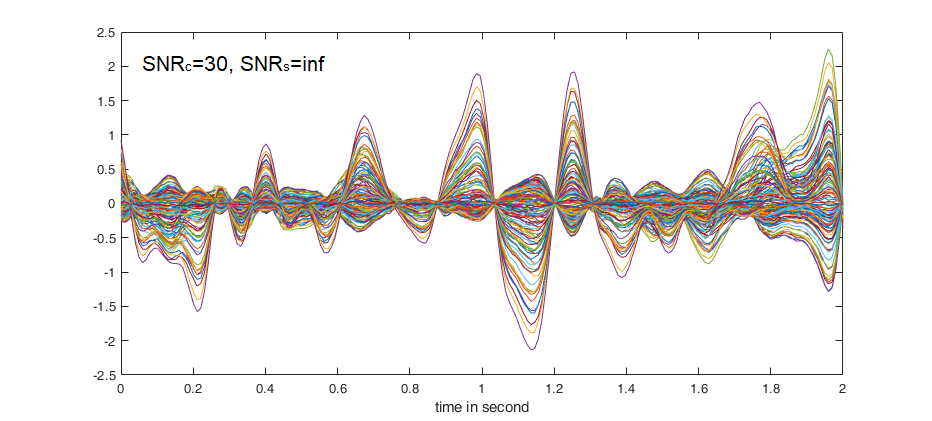}%
	\end{subfigure}
	%	\vspace{0em}
	%	\centering
	\begin{subfigure}[b]{0.8\textwidth} 
		\centering
		\includegraphics[width=0.75\textwidth,height=0.18\textheight]{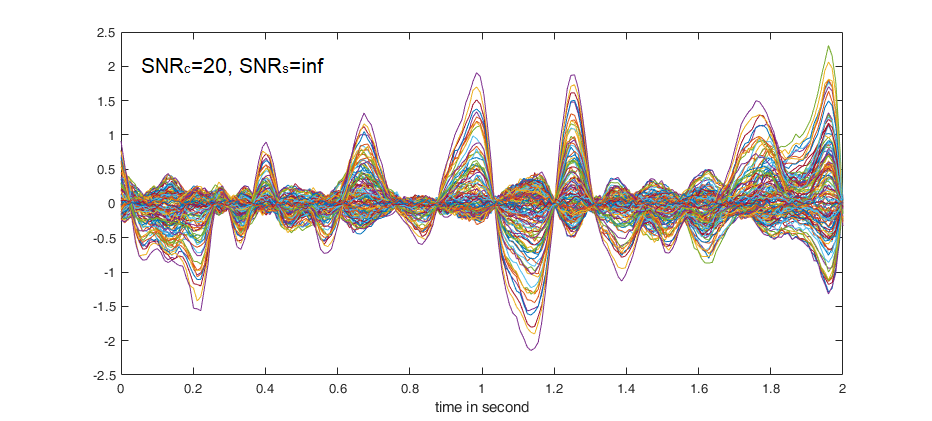}%
	\end{subfigure}
	\begin{subfigure}[b]{0.8\textwidth} 
		\centering
		\includegraphics[width=0.75\textwidth,height=0.18\textheight]{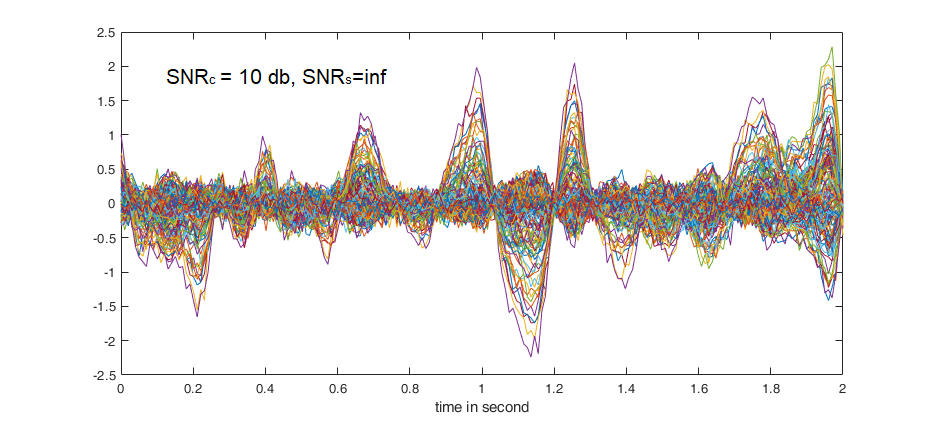}%
	\end{subfigure}
	\caption{Part 1: Impact of noise in electrode channels and brain sources on the EEG signal from at different SNR levels.}	
	\label{fig:noise_impact}		
\end{figure*}

\begin{figure*}[h]	
	\centering	
	\begin{subfigure}[b]{0.8\textwidth} 
		\centering
		\includegraphics[width=0.75\textwidth,height=0.18\textheight]{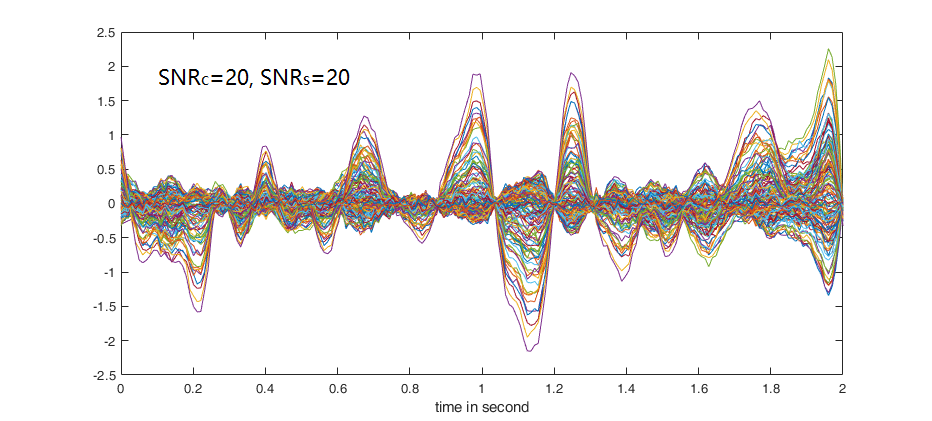}%
	\end{subfigure}
	\begin{subfigure}[b]{0.8\textwidth} 
		\centering
		\includegraphics[width=0.75\textwidth,height=0.18\textheight]{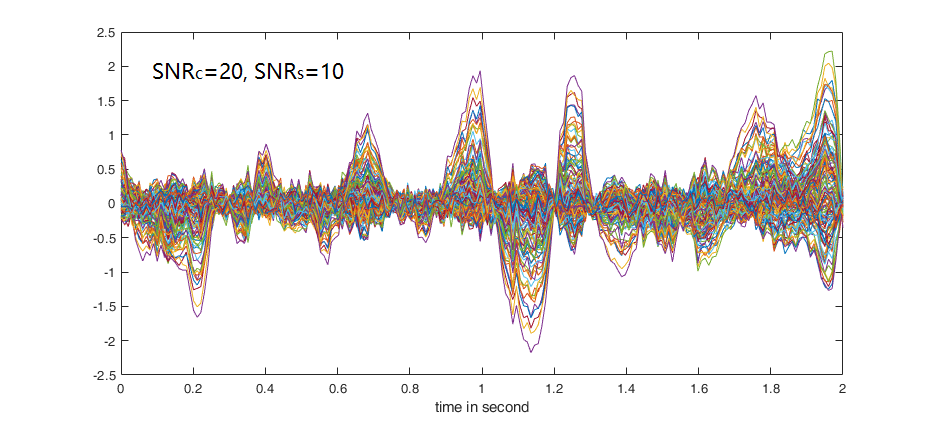}%
	\end{subfigure}
	\begin{subfigure}[b]{0.8\textwidth} 
		\centering
		\includegraphics[width=0.75\textwidth,height=0.18\textheight]{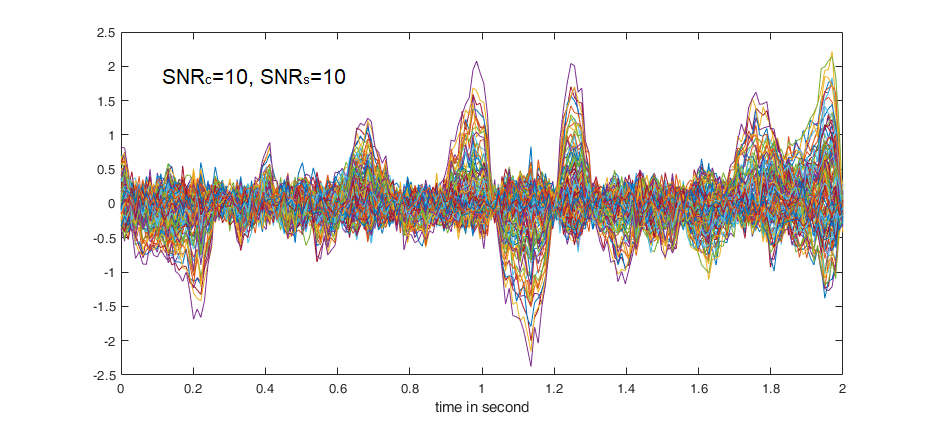}%
	\end{subfigure}
	\caption{Part 2: Impact of noise in electrode channels and brain sources on the EEG signal from at different SNR levels.}
	\label{fig:noise_impact_part2}		
\end{figure*}

\begin{figure*}[h]	
	\centering	
	\begin{subfigure}[b]{0.75\textwidth} 
		\centering
		\includegraphics[width=0.98\textwidth]{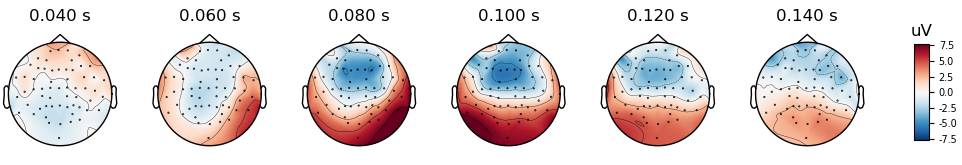}%
	\end{subfigure}
	%	\vspace{5em}		
	\begin{subfigure}[b]{0.75\textwidth} 
		\centering
		\includegraphics[width=0.98\textwidth]{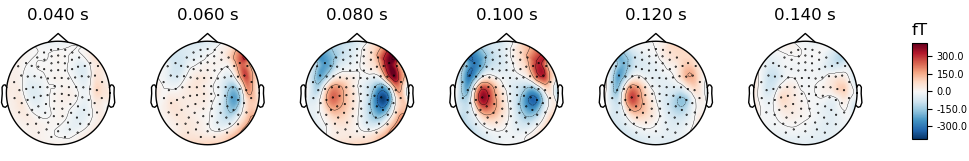}%
	\end{subfigure}
	%	\vspace{0em}
	%	\centering
	\begin{subfigure}[b]{0.75\textwidth} 
		\centering
		\includegraphics[width=0.98\textwidth]{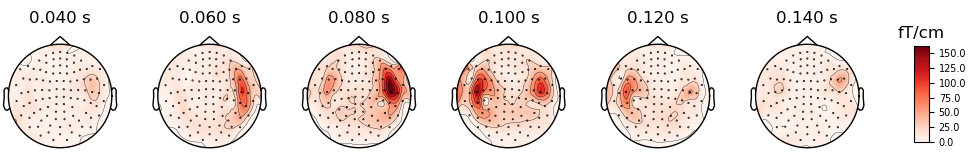}%
	\end{subfigure}
	\caption{Topomaps of EEG (top row) and MEG including mag (middle row) and gradiometers (bottom row) for left auditory stimuli.}	
	\label{topo_realdata}
\end{figure*}

\begin{figure*}[h]
	\centering
	\includegraphics[clip,width=0.65\columnwidth]{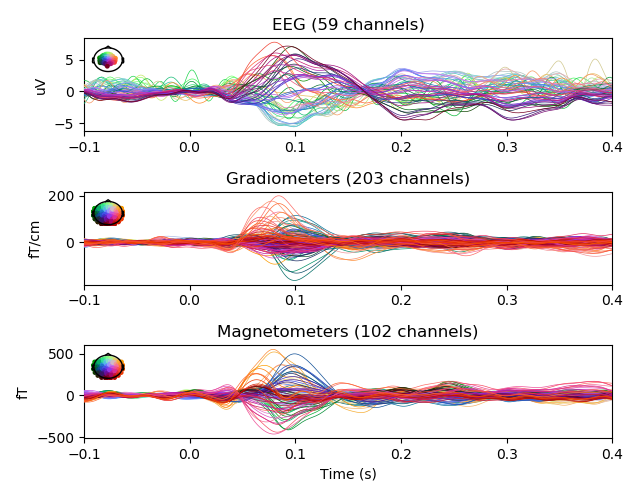}%
	\caption{Time series plots for signal from EEG, Gradiometers, and Magnetometers channels for left auditory stimuli}
	\label{fig:tsplots}
\end{figure*}

\begin{figure*}[h]	
	\centering	
	\begin{subfigure}[b]{0.75\textwidth} 
		\centering
		\includegraphics[width=0.98\textwidth]{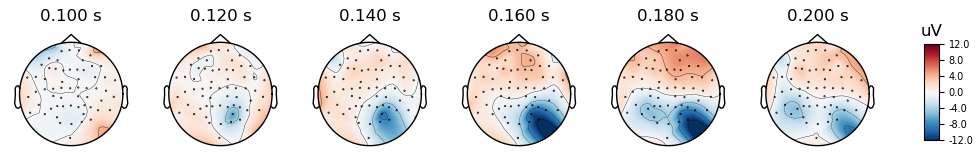}%
	\end{subfigure}
	%	\vspace{5em}		
	\begin{subfigure}[b]{0.75\textwidth} 
		\centering
		\includegraphics[width=0.98\textwidth]{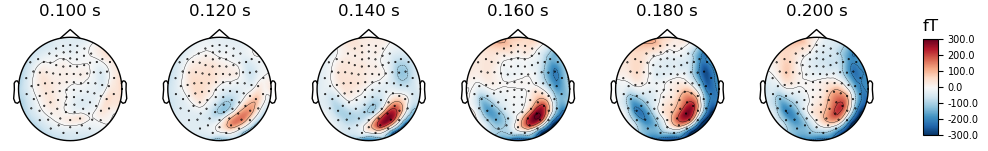}%
	\end{subfigure}
	%	\vspace{0em}
	%	\centering
	\begin{subfigure}[b]{0.75\textwidth} 
		\centering
		\includegraphics[width=0.98\textwidth]{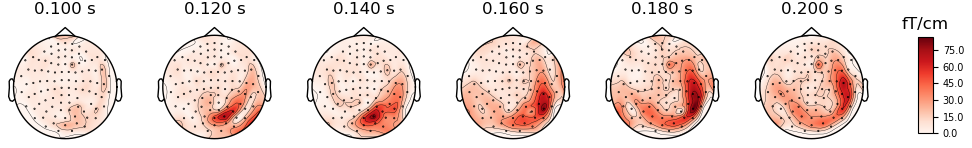}%
	\end{subfigure}
	\caption{Topomaps of EEG (top row) and MEG including mag (middle row) and gradiometers (bottom row) for left visual stimuli.}	
	\label{topo_realdata_left_visual}		
\end{figure*}

\begin{figure*}[h]
	\centering
	\includegraphics[clip,width=0.65\columnwidth]{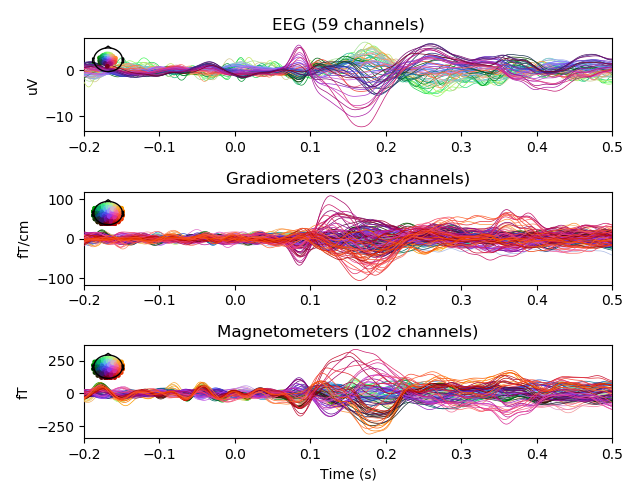}%
	\caption{Time series plots for signal from EEG, Gradiometers, and Magnetometers channels for left visual stimuli}
	\label{fig:tsplots_lv}
\end{figure*}

\end{document}